# Lipid Lateral Diffusion: Mechanisms and Modulators

V. K. Sharma[1,2*], H. Srinivasan[1,2], J. Gupta[1,2], S. Mitra[1,2]

[1]*Solid State Physics Division, Bhabha Atomic Research Centre, Mumbai, 400085, India*

[2]*Homi Bhabha National Institute, Mumbai, 400094, India*

*Abstract*

The lateral diffusion of lipids within membrane is of paramount importance, serving as a central mechanism in numerous physiological processes including cell signaling, membrane trafficking, protein activity regulation, and energy transduction pathways. This review offers a comprehensive overview of lateral lipid diffusion in model biomembrane systems explored through the lens of neutron scattering techniques. We examine diverse models of lateral diffusion and explore the various factors influencing this fundamental process in membrane dynamics. Additionally, we offer a thorough summary of how different membrane-active compounds, including drugs, antioxidants, stimulants, and membrane proteins, affect lipid lateral diffusion. Our analysis unveils the intricate interplay between these additives and membranes, shedding light on their dynamic interactions. We elucidate that this interaction is governed by a complex combination of multiple factors including the physical state and charge of the membrane, the concentration of additives, the molecular architecture of the compounds, and their spatial distribution within the membrane. In conclusion, we briefly discuss the future directions and areas requiring further investigation in the realm of lateral lipid diffusion, highlighting the need to study more realistic membrane systems.

\* Email : sharmavk@barc.gov.in; vksphy@gmail.com  Phone : +91-22-25594604

# INTRODUCTION

A cell serves as the fundamental unit of life, exemplifying a dynamic interplay of complex molecular structures and processes. Central to this complexity is the cell membrane, a natural hydrophobic barrier, which delineate the cytosol from the extracellular environment. The cell membrane is not merely a structural entity but plays pivotal roles in numerous biological processes, encompassing selective permeability, cell defense, recognition, adhesion, and signaling. The fluid mosaic model [1], proposed by Singer & Nicolson in 1972, laid the groundwork for understanding the structural and functional complexity of cell membranes. This model describes each leaflet of the cell membrane as a two-dimensional homogeneous fluid of lipids interspersed with proteins and carbohydrates. Subsequent refinements to this model have incorporated the concept of phase-separated microdomains or rafts within the membrane, each exhibiting distinct composition and dynamics compared to the surrounding fluid phase [2]. The lipid raft hypothesis is a specific interpretation of the broader concept of lateral membrane inhomogeneity. In 2006, a consensus operational definition of 'lipid rafts' was established, based on available evidence suggesting that rafts are heterogeneous and dynamic (in terms of both lateral mobility and association-dissociation) cholesterol and sphingolipid enriched membrane nano-domains (10–200 nm)[3]. These nano-domains have the potential to form larger microscopic domains (>300 nm) when clustering is triggered by protein–protein and protein–lipid interactions. The enrichment of these hydrophobic components gives lipid rafts unique physical properties, such as increased lipid packing and order, along with reduced fluidity. These domains are present in both the inner and outer leaflets of asymmetric cell membranes, are likely coupled across leaflets, and serve as functional platforms for regulating various cellular processes[4, 5]. Recent advancements in membrane biophysics have highlighted the heterogeneous nature of cell membranes, consisting of a dynamic interplay between lipids, proteins, and other small molecules, such as carbohydrates[6]. Figure 1 provides a general overview of lateral heterogeneity in the plasma membrane. Given the pivotal role of lipid rafts in numerous biological processes, their dynamics are intricately governed by the microscopic behavior of lipids. Specifically, the lateral diffusion of lipids within these membranes, propelled by thermal agitation, significantly influences their distinctive viscoelastic characteristics.

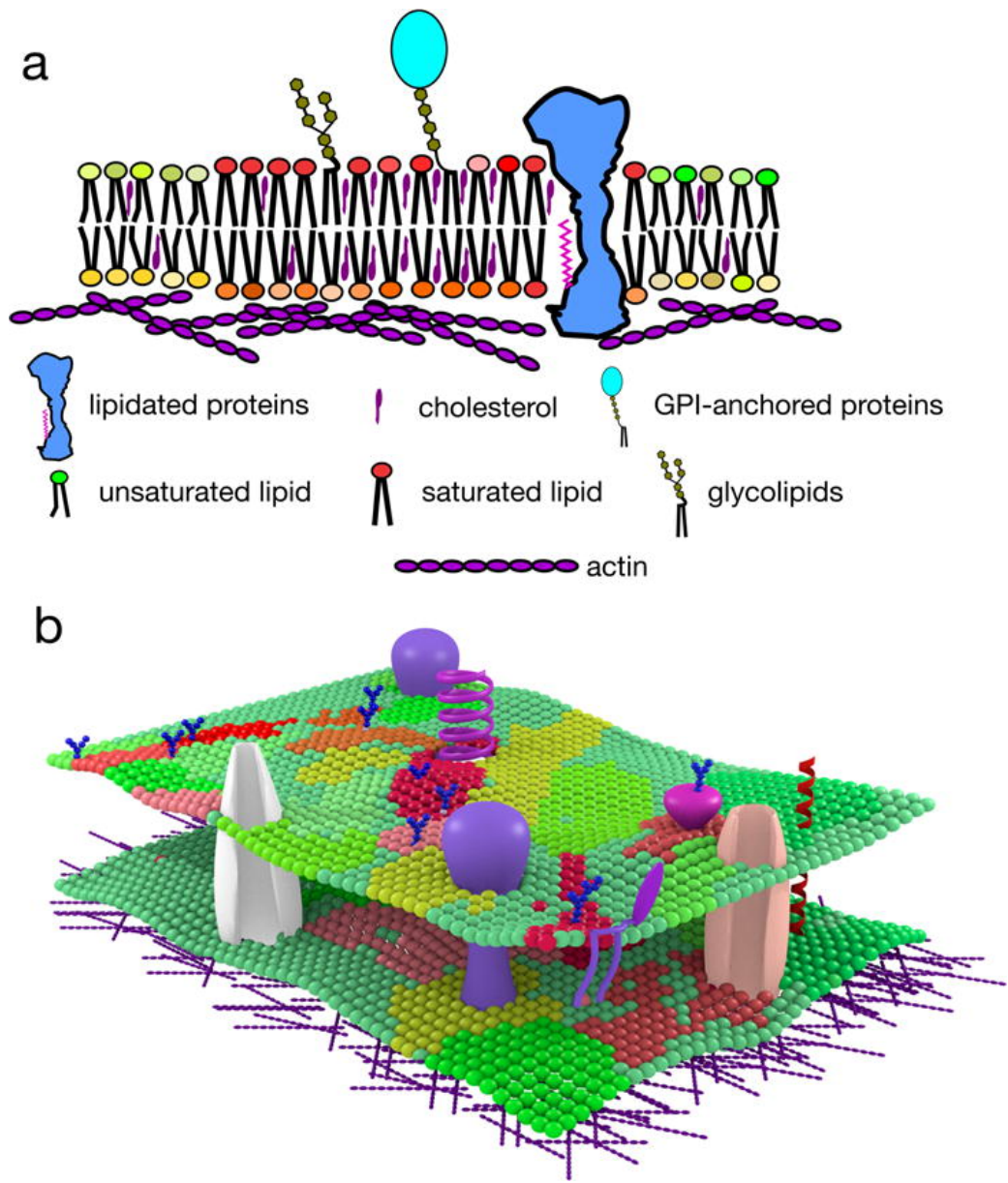

**Fig. 1.** Overview of lateral heterogeneity in the plasma membrane shown in (a) a 2-dimensional view and (b) a 3-dimensional view (adapted from ref. [5])

Despite being only few nanometer thick, cell membranes possess sufficient elasticity to contain the cytoplasm while retaining fluidity to facilitate the movements of lipids and proteins essential for their functionality[7]. The basic matrix of the cell membrane comprises dynamic, fluid self-assemblies of lipids, called lipid bilayer. This lipid bilayer structure serves as the foundation for the assembly of membrane proteins and the formation of specialized microdomains that regulate various cellular processes[8]. In the pursuit of deciphering the intricate biological processes regulated by cell membranes at the molecular level, model membrane systems, such as liposomes and lipid bilayers, have emerged as invaluable tools[9]. These model systems enable researchers to probe membrane dynamics, which significantly influence membrane fluidity and viscoelastic behavior, thereby playing pivotal roles in physiological processes such as cell signaling, membrane trafficking, permeability, vesicle fusion, and endo- or exocytosis[5].

Lipids serve as the foundational components of the lipid bilayer, offering structural integrity and functionality. Lipids are amphiphilic molecules characterized by hydrophilic heads and hydrophobic fatty acid tails. Lipids make up roughly half the mass of most cell membranes, although this ratio can differ depending on the membrane type. For instance, plasma membranes consist of about equal parts lipids and proteins. Bacterial plasma membranes are often composed of one main type of phospholipid and contain no cholesterol. For example, the plasma membrane of gram-negative bacteria (e.g., E. coli) consists predominantly of phosphatidylethanolamine (PE), which constitutes 80 % of total lipids[10]. In contrast, the plasma membrane of gram positive bacteria (e.g. S. *aureus*) is mainly composed of phosphatidylglycerol (PG) lipid, which is about 55 % of total lipids[10]. Mammalian plasma membranes are more complex, containing four major phospholipids—phosphatidylcholine (PC), phosphatidylserine (PS), PE, and sphingomyelin(SM)—which together constitute 50 to 60% of total membrane lipid. In addition to the phospholipids, the plasma membranes of mammalian cells contain glycolipids and cholesterol, which generally correspond to about 40% of the total lipid molecules. The diverse class of lipid molecules exhibits variations in the head group, tail length, and tail unsaturation[11]. Within cellular environments, lipid structures have evolved to fulfill specific functional needs, resulting in varied concentrations across organelle membranes and cellular tissues[6, 12, 13]. Notably, mammalian membranes boast over thousand unique lipid species, each playing distinct roles in biological processes. Among these lipid classes, phospholipids stand out as a major group, featuring a phosphate headgroup and two hydrophobic fatty acid chains linked by an alcohol residue, typically glycerol. The diversity of phospholipids is evident in their various head groups (Fig. 2), with PC being predominant in over half of mammalian plasma membrane phospholipids[14]. By altering head group, alkyl chain length, and tail unsaturation, a plethora of phospholipid variations emerge. For examples, different phospholipids with varying head groups such as PC, phosphatidic acid (PA), PE, PS, PG, and phosphatidylinositol (PI) are shown in Fig. 2. The amphiphilicity of lipids drives their self-assembly into diverse structures such as micelles, vesicles, and bilayers. Additionally, recent research has explored synthetic lipids like dialkyldimethylammonium bromide, which hold promise for applications in drug delivery and gene delivery systems.

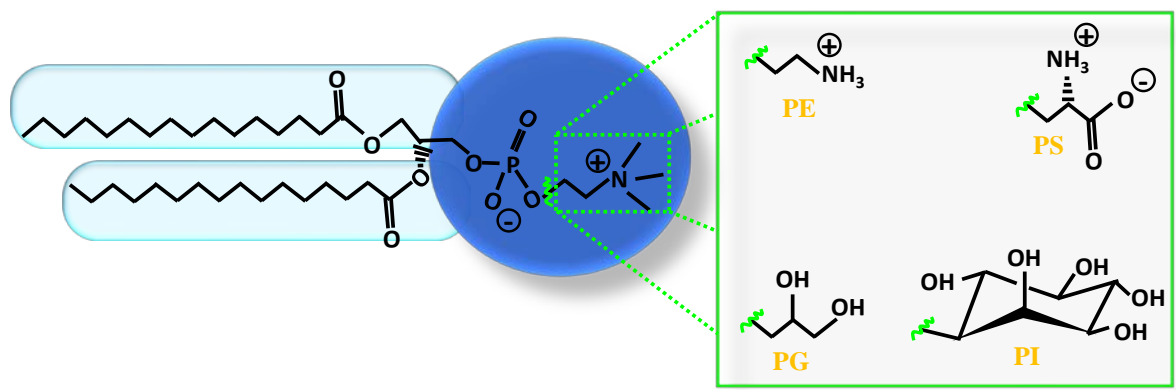

**Fig. 2** Schematic illustrating the chemical structure of a phosphatidylcholine (PC) and other phospholipids featuring diverse headgroups, such as phosphatidylethanolamine (PE), phosphatidylserine (PS), phosphatidylglycerol (PG), and phosphatidylinositol (PI).

Lipid membranes are complex and dynamic entities, displaying a hierarchy of dynamics[15-21]. This spectrum encompasses individual movements of lipids, such as vibrations, lateral diffusion, flip flop, and rotation. Additionally, these membranes exhibit collective behaviors where multiple lipid molecules synchronize their movements, leading to phenomena like membrane bending and fluctuations in membrane thickness. Figure 3 illustrates the schematics of these various molecular and collective motions exhibited by lipid membranes. These diverse dynamical motions cover a broad range of time scales, extending over many decades. They range from rapid molecular vibrations occurring within tens of femtoseconds to slower flip flops that unfold over a few hours. Moreover, the characteristic length scales associated with these motions vary from Angstroms for localized molecular movements to several micrometers for macroscopic cellular deformations. To probe these motions, a variety of spectroscopic methods have been employed, including nuclear magnetic resonance (NMR)[22-25], fluorescence correlation spectroscopy (FCS)[26], dynamic light scattering (DLS)[27, 28], electron paramagnetic resonance (EPR)[29], x-ray photo-correlation spectroscopy (XPCS)[30], neutron spin echo (NSE)[31-36] and quasi-elastic neutron scattering (QENS)[16, 20, 21, 37-46]. However, despite the extensive range of spectroscopic techniques utilized, many are constrained to measuring dynamics within a limited time and length scale. NMR, FCS, and DLS primarily measure diffusion over length scales greater than a micrometer and time scales longer than nanoseconds, making them generally classified as macroscopic methods. On the other hand, QENS probes lipid membrane dynamics on time scales ranging from sub-picoseconds to nanoseconds and length scales spanning from a few Angstroms to nanometers, thus typically falling under the category of microscopic methods. In this length and time scale, lipid molecules perform long range lateral diffusion of lipid within the leaflet and localised internal motion of the lipids[43, 47-51]. QENS have been widely used to study lateral and internal motion of lipids within the membranes[15-17, 20, 37, 40, 43, 51-55]. The spatial-temporal range explored by QENS can be expanded to submicroseconds and several tens of nanometers by utilizing neutron spin echo instruments[20, 40]. NSE has been used to study collective bending and thickness fluctuations of the membranes[15, 24, 34, 56, 57]. The time and length scales accessible through QENS are highlighted in green in Fig. 3. More information about multiscale membrane dynamics can be found from elsewhere[55].

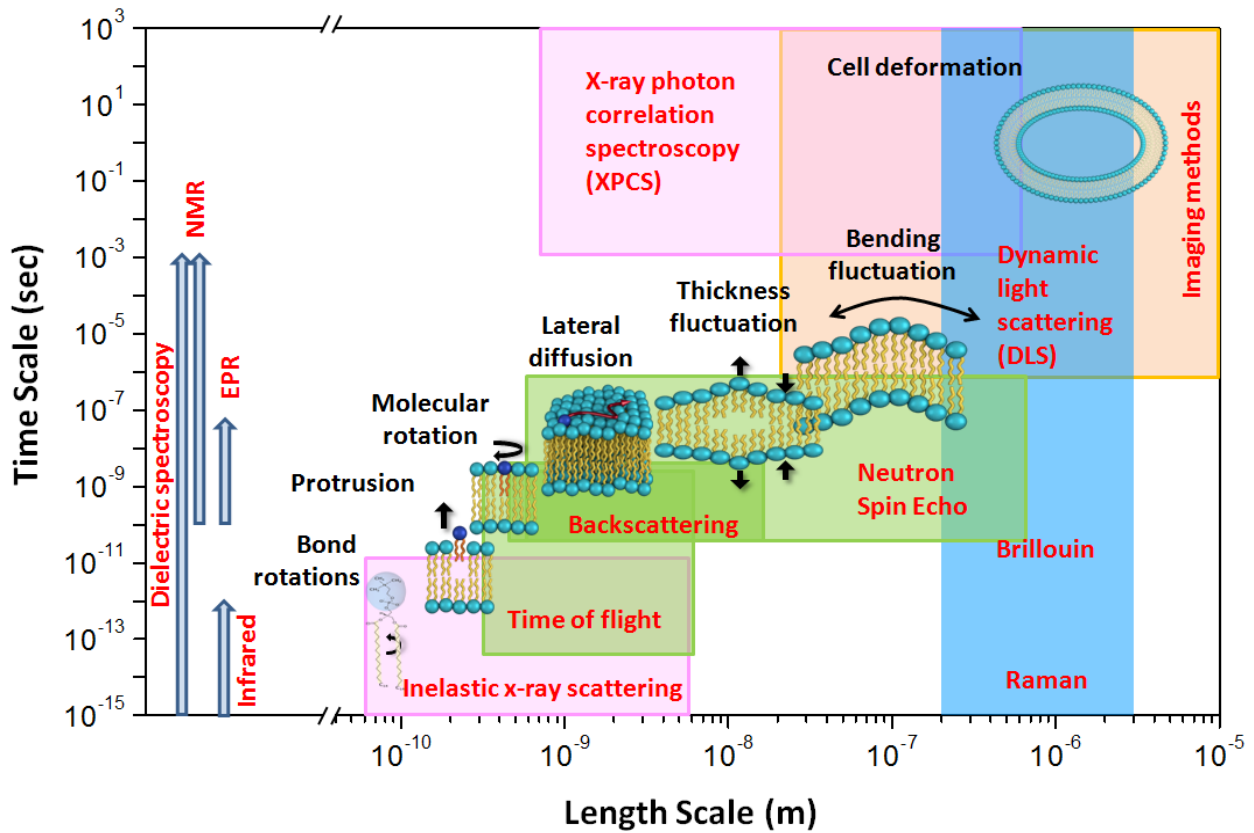

**Fig. 3** Schematic illustrating dynamical processes observed in typical lipid membranes, along with their corresponding temporal and spatial regimes. The length and time scales accessible by various spectroscopic methods are also depicted. For direct comparison, neutron scattering are represented by green squares, x-ray scattering by magenta squares, light scattering by blue squares, and imaging methods by yellow square. Other spectroscopic methods shown on the leftmost side, such as electron paramagnetic resonance, nuclear magnetic resonance, infrared, and dielectric spectroscopy, cover a broad temporal regime without specific spatial associations (adapted from ref.[20]).

In this review article, we will focus on lateral lipid diffusion, which refers to the long-range movement of entire lipid molecules within the membrane's leaflet, driven by thermal agitation. Lateral diffusion plays a crucial role in modulating membrane fluidity, organization, and functionality and has a central role in various physiological processes including cell signaling, membrane trafficking, membrane protein activity, and energy transduction pathways. Lateral diffusion facilitates the creation of transient microdomains, known as rafts, which are enriched in specific lipids. These rafts play a crucial role in signal transduction and the cellular response to stimuli. Additionally, lateral diffusion facilitates the redistribution of lipids to repair damaged regions in the membrane, thereby preserving its integrity. Lateral motion is essential for maintaining the fluidity of the cell membrane, which in turn regulates its transport properties. Membrane fluidity also plays an important role in endocytosis and exocytosis which facilitates material uptake and waste removal by cells. The semi-permeable nature of the cell membrane selectively allows essential elements to permeate into and out of cells while blocking harmful components to enter into the cells. Alterations in lateral lipid diffusion can impact membrane fluidity, affect protein-lipid interactions, and modulate cellular responses to external stimuli. For example, reduced lateral diffusion of lipids leads to decreased fluidity in the cell membrane, potentially impairing its permeability and hindering the transport of vital components. Conversely, accelerated lateral lipid diffusion enhances membrane fluidity, increasing permeability and allowing harmful substances to enter cells, which can cause cellular damage. Therefore, understanding the mechanisms and regulation of lipid lateral diffusion is crucial for elucidating the complex dynamics of cell membranes and their role in health and disease. Despite extensive research efforts, the determination of lateral diffusion coefficients for lipid molecules within membranes remains a topic of debate. The reported values for these coefficients exhibit significant variability, contingent upon the time and length scales probed by different measurement or computational methods[58]. Previous reviews on lateral diffusion, such as those by Vaz, Zalduondo and Jacobson[59] Clegg and Vaz[60], Almeida and Vaz[61], Tocanne[62], and Saxton[63] primarily covered studies up to 1999. In these review articles, lateral diffusions were observed mainly through the macroscopic techniques, predominantly based on fluorescence based techniques such as Fluorescence recovery after photobleaching (FRAP) and single particle tracking (SPT). In 2009, Lindblom and Oradd have reviewed the measurement of lateral diffusion using pulsed field gradient (PFG)-NMR method[58]. In this review article, we concentrate on recent studies concerning lateral lipid diffusion at the nanometer scale, particularly employing microscopic neutron scattering techniques. We explore different models proposed for lateral motion based on QENS experiments and discuss various theories aimed at elucidating lateral diffusion, along with factors influencing this motion. Additionally, we also provide an overview of the impact of various membrane-active compounds on lipid lateral diffusion.

In this article, we begin by exploring various models of lateral motion of lipids discussed in the literature. We then venture into the realm of neutron scattering in particular QENS, a prominent method used to observe lateral motion lipid at the microscopic length scale. In the next section, we delve into the theoretical framework of lateral lipid diffusion and examine the multitude of factors that influence this dynamic process. In the following section, we illuminate the effects of various membrane-active compounds on lateral lipid diffusion. In the last section, we summarize our findings with concluding remarks and highlight promising avenues for future research in this intriguing field.

**DIFFERENT MODELS OF DIFFUSION**

The dynamics of the lipid membrane exhibit a level of complexity beyond of ordinary liquids. In a simple liquid[64], atoms experience minimal interaction with surrounding atoms at short times, resulting in ballistic motion where the mean square displacement (MSD) is proportional to the square of time. This ballistic motion transitions to Fickian diffusion at longer times, characterized by an MSD which is proportional to time. However, in dense liquids, a phenomenon known as caging effect occurs, wherein atoms are temporarily confined by their neighbors, leading to a power-law dependence of MSD on time. In the context of lipid bilayers, the motion of lipid atoms becomes even more intricate due to the high flexibility of the molecules. This flexibility gives rise to a pronounced subdiffusive regime between ballistic and normal Fickian diffusion[65]. In this section, we will discuss in brief about these different models of motions.

**(a) Ballistic motion**

In the case of a ballistic motion, the displacement of the particle ($s$) can be written as a function of time ($t$), simply defined by its velocity ($v$)

$$s = vt \tag{1}$$

In this case, the self Van Hove correlation function, $G(r,t)$, will be a Gaussian in space and can be written as

$$G^{BM}(r,t) = \frac{1}{\left(2\pi v^2 t^2\right)^{3/2}} \exp\left(-\frac{r^2}{2v^2 t^2}\right) \tag{2}$$

Here, BM in superscript is for ballistic motion. The spatial Fourier transform of self Van Hove correlation function gives an incoherent intermediate scattering function (IISF) which is a Gaussian in momentum transfer ($\hbar Q$) as well as in time space

$$I_{inc}^{BM}(Q,t) = \exp\left(-\frac{Q^2 v^2 t^2}{2}\right) \tag{3}$$

Hence, the incoherent scattering law, Fourier transform of Eq. (3) with respect to time will be a Gaussian in the energy transfer ($\hbar\omega$)

$$S_{inc}^{BM}(Q,\omega) = \frac{1}{\sqrt{2\pi} Qv\hbar} \exp\left(-\frac{\omega^2}{2Q^2 v^2}\right) \tag{4}$$

Therefore, for a ballistic motion, scattering law will be Gaussian and the full width at half maximum (FWHM) of this scattering function will be $2.35Qv$ i.e. linear with $Q$. The velocity ($v$) can be obtained from the slope between FWHM and $Q$. If the ballistic motion is primarily influenced by thermal energies and friction is neglected, then the anticipated velocity of the molecule should resemble that of a free particle. Consequently, it is determined by the thermal energy, as prescribed by the equipartition theorem $v = \sqrt{(2k_B T/M)}$, where $k_B$ is Boltzmann constant, T is Temperature and $M$ is the mass of the particle.

**(b) Brownian motion: Continuous diffusion**

For a particle diffusing via random Brownian motion, the displacement of the particle can be written as a function of time,

$$s = \sqrt{2NDt} \tag{5}$$

where $N$ is the dimension, $D$ is translational diffusion coefficient of the particle. In this case, the self Van Hove correlation function can be obtained by solving the Fick's Equation[46] and found to be a Gaussian in space which can be written as

$$G^{CD}(r,t) = \frac{1}{(4Dt)^{N/2}} \exp\left(-\frac{r^2}{4Dt}\right) \tag{6}$$

Here, CD in superscript represents continuous diffusion. The spatial Fourier transform of Eq. (6) gives an IISF which is a Gaussian in $Q$ and exponential decay in time

$$I_{inc}^{CD}(Q,t) = \exp(-DQ^2 t) \tag{7}$$

Its temporal Fourier transform will be a Lorentzian scattering function in energy transfer

$$S_{inc}^{CD}(Q,\omega) = \frac{1}{\pi}\left(\frac{DQ^2}{(DQ^2)^2 + \omega^2}\right) \tag{8}$$

Hence, in case of Brownian motion, quasi-elastic profile will be a Lorentzian with a half width at half maximum (HWHM) that varies quadratically with $Q$, in particular $DQ^2$. Notably, the HWHM of the Lorentzian increases linearly with $Q^2$, indicating continuous diffusion. By analyzing the slope between the HWHM and $Q^2$, one can determine the diffusion coefficient.

**(c) Jump Diffusion**

In systems where the interactions between particles are strong, often the particles experience transient caging and sudden jumps. These intermittent jumps can be described through jump-diffusion process, wherein the particle makes sudden jumps interspersed by particle being caged for an average residence time, τ. Modeling the jump-diffusion involves considering the aspect of caging and jumping which occur alternatively during the process. During the caging, the particles undergo caged diffusion, which essentially can be treated as diffusion within a confined spherical volume with reflecting boundary conditions. In this scenario, the displacement of the particle can be written as

$$s = \sqrt{6\left(b^2\left[1 - e^{-\gamma t}\right] + \left(\frac{l_0^2}{\tau}\right)t\right)}$$

where the $b$ is the radius of confinement, $\gamma$ denotes the rate of caged of diffusion, and $l_0$ is the root mean-squared jump-length. While, the MSD or displacement of the particle can be explicitly given, it should be noted that the jump-diffusion process is strongly non-Gaussian in nature and therefore the MSD doesn't contain sufficient information about the nature of the dynamics. This is because, higher moments of displacement (higher than 2), start contributing strongly to the self van Hove correlation function. In fact, a closed form expression for the cage-jump diffusion model doesn't exist for self van Hove correlation function. However, at times sufficiently longer than the individual caging times, the jump diffusion can be modeled independently using continuous time random walk (CTRW) models. In this model, the particle executes jumps of certain length, $l$, sampled from jump-length distribution, $\rho(l)$. The solutions to the CTRW model can be conveniently obtained in Fourier space, in terms of IISF,

$$I_{inc}^{JD}(Q,t) = \exp\left[(1-\rho(Q))\frac{t}{\tau}\right] \tag{9}$$

where, JD in superscript represents jump diffusion, $\rho(Q)$ is the Fourier transform of the jump-length distribution and $\tau$ is the mean residence between subsequent jumps. Rewriting $\tau^{-1}[\rho(Q) - 1] = \Gamma_j(Q)$, we get

$$I_{inc}^{JD}(Q,t) = \exp\left[-\Gamma_j(Q)t\right] \tag{10}$$

indicating that even for the jump-diffusion process, the intermediate scattering function is an exponential decay in time. Therefore, the QENS spectra for a jump-diffusion process will also follow a Lorentzian profile of the form,

$$S_{inc}^{JD}(Q,\omega) = \frac{1}{\pi}\frac{\Gamma_j(Q)}{\Gamma_j(Q)^2 + \omega^2} \tag{11}$$

where $\Gamma_j(Q)$ is the HWHM of the Lorentzian and it's $Q$-dependence is dictated by the nature of jump-length distribution. Among the most common models, considering a radial symmetric exponential jump-length distribution gives a standard model for $\Gamma_j(Q)$,

$$\Gamma_j(Q) = \frac{D_j Q^2}{1 + \tau D_j Q^2} \tag{12}$$

where $D_j$ is the jump-diffusivity and is defined as $D_j = (l_0)^2/(6\tau)$. Notably, in the limit $Q \to 0$, which reflects the behavior at long distances, the HWHM of the jump-diffusion goes into the limit of Brownian motion, $\Gamma_j(Q) = D_j Q^2$.

**(d) Confined diffusion**

As discussed in the previous example of cage-jump diffusion, the confined diffusion model can be described considering the diffusion within a spherical volume of radius $b$, wherein the spherical walls can be regarded to have reflecting boundary conditions. In this scenario, the displacement is essentially given by,

$$s = b\sqrt{6\left[1 - e^{-\gamma t}\right]} \tag{13}$$

Evidently, at long times $t \to \infty$ the displacement in this model saturates to a constant value $b\sqrt{6}$. However, it is important to note that the displacement doesn't contain complete information about the nature of the diffusion process. In the Fourier space, a direct expression for the IISF can be given as a sum of decaying exponential and a constant,

$$I_{inc}^{CnD}(Q,t) = A_0(Qb) + \left[1 - A_0(Qb)\right]e^{-\Gamma_{loc}(Q)t} \tag{14}$$

where CnD in superscript represents confined diffusion, $A_0(Qb)$ is elastic incoherent structure factor (EISF) which contains information about the geometry of the confinement region. Typically, the exact analytical expression for localized diffusion process will be denoted as a series of exponentials, but here we have considered it to be approximated to a single exponential decay, with a decay constant $\Gamma_{loc}$, which refers to the localized nature of the confined diffusion process. The expression for IISF can be used for obtaining the typical QENS spectra,

$$S_{inc}^{CnD}(Q,\omega) = A_0(Qb)\delta(\omega) + \left[1 - A_0(Qb)\right]\frac{1}{\pi}\frac{\Gamma_{loc}(Q)}{\Gamma_{loc}(Q)^2 + \omega^2} \tag{15}$$

where the delta function contributes to an elastic peak in the QENS spectra. The coefficient of the elastic peak, $A_0(Qb)$, is the EISF and it strongly depends on the value of radius of confinement $b$ and momentum transfer $Q$. Although an analytic expression for the behaviour $\Gamma_{loc}(Q)$ doesn't exist, it can be computed numerically based on the model of localized diffusion within a sphere.

### (e) Anomalous Diffusion

In case of anomalous diffusion, displacement of particle can be written as

$$s = \sqrt{At^\alpha} \tag{16}$$

where α is the sub-diffusion exponent which can take a value, $0 < \alpha < 1$ and $A$ is associated with the sub- diffusion constant. In the case of lipid membrane, the sub-diffusive motion of lipids in the bilayer can be associated with crowding of lipids in the system, which can be modeled as a non-Markovian[66] diffusion process. In this scenario, the non-Markovianity or history dependence is linked to the fact that lipids undergo numerous backscattering events during the diffusion process. In this case, the self Van Hove correlation function for diffusion in 3D can be given by,

$$G^{AD}(r,t) = \frac{1}{\left(4\pi At^\alpha\right)^{3/2}}\exp\left[-\frac{r^2}{4At^\alpha}\right] \tag{17}$$

This essentially results an IISF which is a Gaussian in $Q$ and a stretched exponential decay in time

$$I_{inc}^{AD}(Q,t) = \exp\left[-AQ^2 t^\alpha\right] \tag{18}$$

An analytic expression for the scattering law doesn't exist, as the Fourier transform of stretched exponential doesn't have a closed-form solution. Nonetheless, in principle, the scattering law can be written as:

$$S_{inc}^{AD}(Q,\omega) = \int_{-\infty}^{\infty} dt\, e^{-i\omega t} \exp\left[-AQ^2 t^\alpha\right] \qquad (19)$$

which can be directly evaluated numerically to fit the obtained experimental data. On the other hand, the data can also be modeled by computing the Fourier transform of experimental $S_{inc}(Q,\omega)$, numerically and model it using the analytical form given for $I_{inc}(Q,t)$.

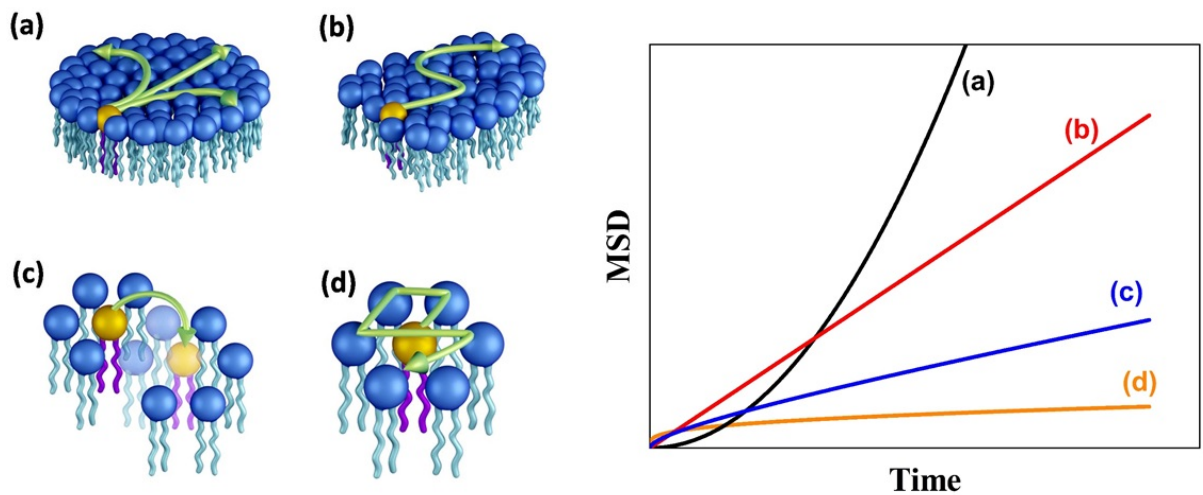

**Fig. 4** (Left) Different mechanisms for the lateral diffusion of lipid: (a) flow like ballistic motion [51, 67], (b) continuous diffusion[47, 68], (c) jump diffusion[69], and (d) confined diffusion[54] (adapted from ref.[55]). (Right) Variation of mean square displacement (MSD) vs time corresponding to each of these models.

A schematic of different models of diffusion and ballistic motion and corresponding variation of MSD with time are shown in Fig. 4. It is evident that these different models can be quantitatively differentiated in traditional plot of MSD with time. For example, ballistic motion will give MSD proportional to square of time but continuous diffusion will give MSD proportional to time. In contrast, in anomalous diffusion, MSD will follow $t^\alpha$, where α is a non-integer value between 0 and 1 and typically takes a value of 0.6 for a variety of lipid bilayer systems[65, 66]. For jump diffusion, at intermediate time, MSD will be sub diffusive, but at long time, it becomes diffusive, i.e., MSD proportional to time. Similarly, dynamic structure factor, $S(Q, \omega)$, varies differently with $Q$ and $\omega$ for these different motions as evident from the above equations. For example, $S(Q, \omega)$ will be a Gaussian, Lorentzian, and Fourier Transform of stretched exponentials for the ballistic motion, continuous diffusion, and anomalous diffusion, respectively. This $S(Q, \omega)$ can be directly observable with the QENS technique which enables one to identify exact diffusion mechanism of particles.

**THEORETICAL FRAMEWORKS FOR LATERAL DIFFUSION**

Several theories have been developed to relate the lateral diffusion coefficient to the microscopic properties of the diffusing molecule and the membrane. In the case of planar lipid membranes, two distinct cases are distinguished according to the size of the diffusing molecules with respect to the size of lipids (which are the basic building blocks of the membrane). The first is a continuum hydrodynamic model[70] for diffusion of particles, the size of which is much larger than that of the lipid. This model is thus best applicable to diffusion of integral membrane proteins in lipid bilayers. The second is free volume model [71, 72] which consider the discreteness of membrane and thus best suited to explain the diffusion of lipid or molecules similar in size to lipids. Both of these models are for the homogenous lipid membrane system. Here, we will briefly discuss both models as well as some additional models for heterogeneous membrane systems.

**(i) Continuum Fluid Hydrodynamic Model**

This model was proposed by Saffman[70, 73] for the diffusion of cylindrical particles within thin, viscous fluid sheets like membranes. This model considers the lipid bilayer as a continuous medium, disregarding both the discrete structure of the membrane and the molecular arrangement of lipids. It conceptualizes the lipid bilayer as a two-dimensional fluid interacting with a bulk three-dimensional fluid, symbolizing the solvent enveloping the membrane on both sides. The initial insight into factors affecting lateral diffusion can be gleaned from the Saffman-Delbrück continuum hydrodynamic model[70]. This model describes lateral and rotational diffusion of an object moving in a two-dimensional fluid, e.g., a lipid membrane. The diffusion coefficient of a cylindrical object of radius $R$ in a membrane with thickness $h$ and viscosity surrounded by bulk liquid with viscosity $\eta_s$ is

$$D_{lat} = \frac{k_B T}{4\pi \eta_m h} \left( \ln\left(\frac{2L_{sd}}{R}\right) - \gamma' \right) \qquad (20)$$

where $\eta_m$ is viscosity of the membrane, and $\gamma'$ is Euler's constant which value ~ 0.577. Here, $L_{sd} = h\eta_m / 2\eta_s$ is the Saffman-Delbrück length scale. It is evident that according to this model, $D_{lat}$ depends on the size of the object and height of the diffusing species in the membrane, viscosity of membrane and surrounded liquid and temperature. The central features of this model include a weak, logarithmic relationship with particle size and a strong dependency on particle height. The basic assumption of continuum hydrodynamic model is that particles composing the fluid are small with respect to the size of diffusing object. However, this condition is not satisfied in the case of lateral diffusion of lipid. Hence, it is not surprising that this model could not explain the lateral diffusion of lipid which has the similar size of lipid composing the membrane. Vaz et al[72] have shown that hydrodynamic model fails to explain the variation of lateral diffusion coefficient of various lipid with varied alkyl chain lengths. It may be noted that diffusion of large size object such as large integral proteins in lipid bilayer is adequately described by Saffeman equations[74]

**(ii) Free Volume model**

This model is based upon the free volume model proposed by Cohen & Turnbull [71] for diffusion in glasses and extended by Galla et al. [75] to describe diffusion in the plane of a membrane. According to free volume model[71, 76], diffusion is limited by the occurrence of a free volume greater than a critical size next to a diffusing particle. Free volumes smaller than the critical size do not contribute to diffusion. In case of a particle performing a random walk in two dimensions, at each diffusion step, the particle needs to have both certain activation energy and a minimum free area to move into. In accordance with Macedo and Litovitz[76], the diffusion coefficient in 2D surface can be written as

$$D_{lat} = D' p(a) p(E) \qquad (21)$$

where $p(a)$ is the probability that the diffusing lipid will find a vacancy next to it of an area greater than a certain critical size and $p(E)$ is the probability that enough energy will be available at each diffusion step to overcome the interactions with neighboring molecules. These probabilities are given by

$$p(a) = \exp\left(-\frac{a_0}{a(T) - a_0}\right)$$

$$p(E) = \exp\left(-\frac{E_a}{k_B T}\right)$$

where $a(T)$ is the average area per lipid, which is a function of temperature, and $a_0$ is the critical area which is essentially the close-packed cross-sectional molecular area. The average free area per molecule in the plane of the bilayer is $a_f = a(T) - a_0$. $E_a$ is the activation energy associated with diffusion. $D'$ can be obtained using the particle performing a random walk in a two dimensional lattice and Eq. (21) can be written as

$$D_{lat} = B\sqrt{\frac{Ta(T)}{M}} \exp\left(\frac{-a_0}{a(T)-a_0} - \frac{E_a}{k_B T}\right) \qquad (22)$$

where B is constant. It is evident that lateral diffusion is influenced by temperature, the free area available in the lipid bilayer, and activation energy. The free area is a characteristic of the bilayer as a whole, and its redistribution plays a crucial role in diffusion. Activation energy indicates the interactions of a lipid molecule with its environment, including neighboring lipids in the bilayer and the surrounding aqueous phase. Evidence supporting the free volume model for lipid diffusion comes from studies showing that the lateral diffusion coefficient of lipids is independent of chain length, as observed by Balcom and Petersen[77] and Vaz et al.[72]. This finding is consistent with the predictions of the free volume model and contradicts the hydrodynamic model.

**(iii) Extended free volume model:**

The free volume model was further extended by including the influence of viscous force in the aqueous phase (due to its contact with water at aqueous interface) and at the mid plane of the bilayer. Vaz et al[72] have shown that in this model, diffusion coefficient is given by

$$D_{lat} = \left(\frac{k_B T}{f}\right) exp\left[\frac{-\zeta a_0}{\{a(T)(\beta + a_a(T-T_m))\}}\right] \qquad (23)$$

where $\zeta$ is a numerical factor that accounts for the overlap of free area ($\zeta$ has values between 0.5 and 1.0), $a_a$ is the lateral thermal expansion coefficient in the fluid phase, and $f$ is the translational friction coefficient resulting from drag force at the membrane-water interface and bilayer mid plane. This $f$ can be written as $f=f_1+f_2$ where $f_1$ is due to interaction of lipid polar head group with aqueous phase at bilayer-water interface and $f_2$ is due to interaction of acyl chain ends of lipid with other half of the bilayer. $f_1$ depends on the viscosity of solvent. $T_m$ is the main phase transition of the lipid

**(iv) Restricted diffusion or Obstruction effects**:

In a homogeneous membrane system or at zero obstacle concentration, the diffusion of species follows either free volume model or continuous hydrodynamic model depending on the size of the species. However, in reality, the cell membrane is a complex heterogeneous mixture of lipids and proteins. Within this heterogeneous environment, obstacles like integral membrane proteins and domain of gel-phase lipids can impede or indirectly influence lipid dynamics, creating barriers to lateral diffusion. Lipid molecules cannot move across these integral protein or solid domains. Furthermore, these obstacles may decrease the available free volume in the lipid phase, leading to a significant reduction in the diffusion coefficient. Consequently, due to obstruction, the lateral diffusion coefficient can be expressed as a product of multiple factors[61, 63]

$$D_{lat}(c) = D_0 * D_{fv}(c) * D_{obst}(c) * D_{hydro}(c) \qquad (24)$$

where $c$ is the area fraction of obstacles, $D_o$ is the diffusion coefficient at zero obstacle concentration, $D_{fv}(c)$, $D_{obst}(c)$ accounts for the direct effect of obstruction, and $D_{hydro}(c)$ accounts for hydrodynamic interactions which are normalized to 1 at $c = 0$. The scope of this review does not extend to other theories (e.g. percolation theory, etc.) which are suitable for heterogeneous membrane systems. More information on these additional theories can be found elsewhere[63].

**METHODS**

**Quasielastic Neutron Scattering**

Thermal and cold neutrons have wavelengths of the order of Å and energies are the order of meV which align closely with the inter-atomic or intra-molecular spacing's, as well as the excitation energies within materials. This inherent compatibility renders neutron scattering an effective method for investigating the structure and dynamics of atoms or molecules in condensed matter. The dynamics within materials typically fall into two broad categories: (i) periodic motions, such as atomic vibrations, and (ii) stochastic motions, including diffusion. These motions typically contribute to distinct ranges of energy transfer. For instance, periodic motions associated with oscillations at a frequency $\omega_0$ result in inelastic peaks at energy transfers $E = \pm \hbar\omega_0$. Conversely, stochastic motions produce signal broadening near the elastic line position ($E=0$), known as quasielastic broadening, which is inversely proportional to the time scale of the motion under scrutiny. Therefore, to observe stochastic motions, measurements are focused on the relatively low energy transfer range (a few µeV to a few meV), centered at the elastic line. In general, this spectrum consists of an elastic and quasielastic contribution. The fraction of the elastic component in the total QENS signal, known as EISF, offers insights into the geometry of molecular motions. Through QENS, both qualitative and quantitative information can be extracted. Qualitative information pertains to the geometric mechanism of the motion, whereas quantitative information relates to the correlation times and length scales of the motion. In the scattering experiments, the intensity of scattered neutrons can be expressed in terms of the double differential scattering cross-section. This cross-section encompasses two distinct contributions arising from coherent and incoherent scattering from the sample, and can be mathematically represented as[46].

$$\frac{d^2\sigma}{dEd\Omega} \propto \frac{k_f}{k_i}[\sigma_{\text{coh}}S_{\text{coh}}(Q,\omega) + \sigma_{\text{inc}}S_{\text{inc}}(Q,\omega)] \quad (25)$$

where, $S_{coh}$ and $S_{inc}$ are the coherent and incoherent scattering laws, $\sigma_{coh}$ and $\sigma_{inc}$ are the coherent and incoherent scattering cross sections; $\omega = \omega_i - \omega_f$ is the energy transfer, and $\mathbf{Q} = \mathbf{k}_i - \mathbf{k}_f$, is the wave vector transfer resulting from the scattering process. While coherent scattering provides insights into pair-correlations within a system, the incoherent scattering component directly reflects the self-correlation function. Notably, the incoherent scattering cross-section of hydrogen ($\sigma_{inc}^H$) is exceptionally high compared to coherent/incoherent scattering cross section of any other atoms ($\sigma_{inc}^H >> \sigma_{inc/coh}^{any\ atom}$). Consequently, in hydrogenous systems, such as lipid membranes, the scattering intensity is predominantly influenced by the incoherent scattering from hydrogen atoms. In these instances, the observed scattered intensity is primarily attributable to the incoherent scattering function, $S_{\text{inc}}(\mathbf{Q}, \omega)$, which represents the space-time Fourier transform of the Van Hove self-correlation function, $G(\mathbf{r},t)$. The Van Hove self-correlation function quantifies the probability of finding a particle at position $\mathbf{r}$ after a time $t$, given its initial position at the origin, $\mathbf{r} = 0$, at time $t = 0$. Therefore, QENS data obtained from hydrogenous systems typically provides valuable insights into the self-diffusion of particles within the system.

For studies of lipid membrane dynamics, $D_2O$ is generally chosen as the aqueous medium instead of $H_2O$. This substitution enhances the scattering contribution from the lipid because deuterium possesses an incoherent scattering cross-section that is 40 times lower than that of hydrogen. Moreover, in the case of vesicle solutions, the solvent contribution can be assessed by measuring QENS spectra from $D_2O$ alone. By comparing the spectra obtained from $D_2O$ alone with those obtained from the vesicle solution in $D_2O$, the scattering signal from the lipid membrane can be isolated. This is achieved by subtracting the solvent contribution from the overall scattering signal.

$$S_{mem}(Q,\omega) = S_{solution}(Q,\omega) - \phi S_{solvent}(Q,\omega) \qquad (26)$$

where $S_{mem}(Q, \omega)$, $S_{solution}(Q, \omega)$ and $S_{solvent}(Q, \omega)$ are the scattering functions for lipid membrane, vesicles solution and solvent, respectively. The factor, $\phi$, takes into account the volume fraction of solvent ($D_2O$) in the solution.

The resultant solvent subtracted spectra, $S_{mem}(Q, \omega)$, is used for data analysis to investigate the diffusion mechanism of the lipids. As mentioned earlier, in the spatial and temporal regime probed by QENS, lipid molecules perform two distinct motions (i) lateral motion of the whole lipid within the leaflet and (ii) localized internal motions of the lipids[43, 48, 55, 78, 79]. While both these motions are stochastic in nature and arise due to thermal fluctuations, their relaxation timescales are well separated. Hence, one can assume these motions are independent to each other. The model scattering function of the membrane can be written as,

$$S_{mem}(Q,\omega) = S_{lat}(Q,\omega) \otimes S_{int}(Q,\omega) \qquad (27)$$

where, $S_{lat}(Q, \omega)$ and $S_{int}(Q, \omega)$ are the scattering functions of the lateral and internal motions of lipid, respectively. Internal motion is localized in nature, hence $S_{int}(Q, \omega)$ can be approximated by a sum of a delta and a Lorentzian function. Lateral motion is highly debatable and various models have been proposed which has corresponding different scattering laws[47, 51, 66, 68]. These scattering laws are employed to fit the observed QENS data to identify the nature of the lateral motion. QENS data at very high energy resolution and at low $Q$ values is highly helpful to identify the mechanism of diffusion as only slower lateral motion will dominant in the observed spectra, faster internal motion will be as background.

**Neutron Elastic Intensity Scan**

A useful measurement in QENS experiment is the Elastic Fixed Window Scan (EFWS) also known as neutron elastic intensity scan. This technique involves monitoring the scattering intensity at zero energy transfer ($E$=0), within the instrumental resolution, as a function of temperature. Any microscopic mobility within the sample causes a shift in the scattering signal intensity away from zero energy transfer, rendering the elastic intensity sensitive to the microscopic dynamics. In cases of purely incoherent scattering, the elastic intensity serves as a measure of the extent of dynamics within the system. Moreover, if the system experiences a phase transition accompanied by a change in its dynamical behavior, this transition manifests as a sudden or discontinuous change in the elastic intensity. Hence, the EFWS serves as a powerful tool for observing phase transitions associated with dynamical changes in the system. It is observed that many phase transitions in lipid membranes, such as pre-transition, main phase transitions, induce significant and sharp changes in neutron elastic scattering intensity, readily detectable through EFWS. Therefore, sudden alterations in elastic scattering intensity suggest that these phase transitions are linked to changes in the microscopic dynamics of lipid membranes, which can be meticulously studied using QENS.

**MECHANISMS OF LATERAL DIFFUSION**

Various techniques have been utilized to explore lateral lipid motion, encompassing both macroscopic methods such as fluorescence-based techniques, PFG-NMR and microscopic methods like QENS. While macroscopic techniques such as fluorescence spectroscopy excel at capturing dynamics occurring over micrometer length scales, QENS offers insights into diffusion over a few lipid diameters (~ 20 Å). The resolution of optical techniques is inherently limited by the wavelength of light, thus predominantly sensitive to dynamics occurring at micrometer length scales. Conversely, in QENS, the largest accessible length scale is determined by the smallest possible $Q$ value, which is contingent on the beam size and beam divergence at small scattering angles. For backscattering/time of flight neutron spectrometers, accessible length scales can extend up to approximately 20 Å. Although the dynamics of lipid molecules have been extensively investigated, a unified model that consistently describes lipid motion across nanometer to micrometer distances is lacking. Additionally, the diffusion constants determined by these two methods often differ significantly, typically by two to three orders of magnitude. For instance, lateral diffusion coefficients measured by macroscopic fluorescence techniques are approximately on the order of $10^{-8}$ cm²/s, whereas those from microscopic QENS techniques are around $10^{-6}$ cm²/s. Different models are employed to explain lipid diffusion in these two experimental approaches. For example, the free volume theory[71, 80] which originates from glass physics to membranes established that lipid molecules undergo a rattling motion within a cage formed by surrounding lipid molecules at short timescales (on the order of a few picoseconds). However, over longer timescales, thermal fluctuations create a void or free volume of similar size to a lipid molecule, enabling the lipid molecule to engage in long-range jump diffusive motion from one void to another. Macroscopic fluorescence techniques measure such long range diffusive motion whereas microscopic QENS measures rattling motion of lipid within the cage of surrounding lipid molecules. For years, the widely accepted free volume theory[80] effectively reconciled differences in diffusion coefficients measured through microscopic and macroscopic methods.

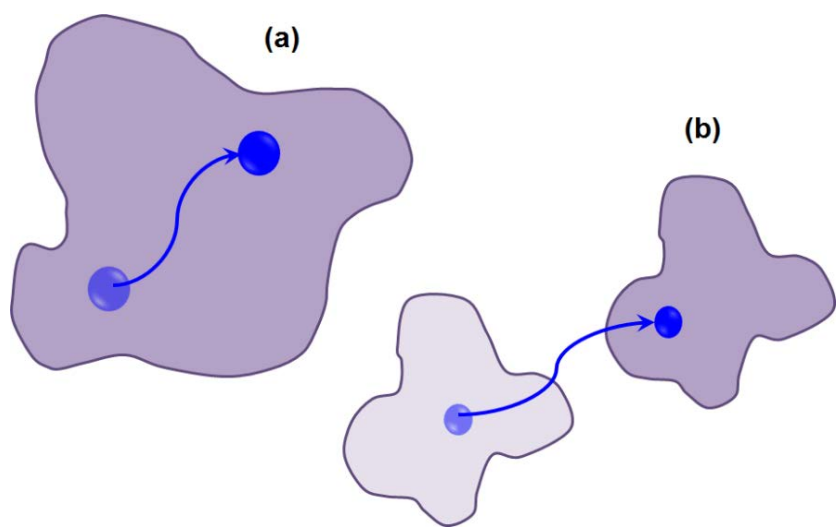

**Fig.** 5 Schematic diagram illustrating two distinct diffusion processes in a lipid bilayer system. (a) The faster diffusion process, where individual lipids move within a confined domain. (b) The slower diffusion process, where lipids move as part of a larger domain.

However, concerns regarding the straightforward mechanistic interpretation of the free volume theory arose upon the discovery, through molecular dynamics (MD) simulations and experimental investigations, that molecular jumps into vacancies were not observed[67, 81]. This is mostly due to the fact that free volume theory was originally developed for colloid like system and basic assumption was that system were composed of hard sphere, which may not be true for lipid as cross-sectional area of lipids is not constant[82]. Instead, in jump like motions, neighbouring lipid molecules were found to exhibit highly correlated motions over distances extending up to the nanometre range[67, 81]. Rheinstädter et al.[81, 83] have proposed two components of lateral diffusion of lipids (i) coherent movement of loosely bound clusters of lipid molecules, which have a diameter of approximately 30 Å and (ii) diffusion of individual lipid within the cluster as shown in Fig. 5. The diffusion of such lipid clusters (Fig. 5 (b)) is slower compared to that of individual lipids within the cluster (Fig. 5(a)). Consequently, it is conceivable that fluorescence measurements are sensitive to the slower motion of a lipid as part of a coherent cluster over long length scales, while the faster, short length scale dynamics of individual lipids are more effectively probed by neutron scattering experiments. These observations indicate presence of at least two pertinent length or time scales associated with the lateral diffusion of lipids in a bilayer.

Using MD simulation, Flenner et al.[65] have shown an extended subdiffusive region in MSD lying between the short time ballistic and longtime Fickian diffusion regimes. Within the subdiffusion model, the lateral diffusivity of the lipid is not a constant but varies as a function of observation timescale. Therefore, they proposed that this extended subdiffusive region is a key factor contributing to the observed disparity in diffusion coefficients obtained from macroscopic and microscopic techniques. Not only do macroscopic and microscopic techniques have different observation length scales, but they also have distinct observation time scales. Microscopic QENS techniques are highly sensitive to sub nanosecond to picosecond time scales, whereas macroscopic techniques are sensitive to time scales on the order of tens or hundreds of nanoseconds. They proposed that due to extended regime, lateral diffusion coefficient obtained through QENS is higher by magnitudes than obtained from macroscopic fluorescence techniques. The primary challenge in reconciling macroscopic and microscopic experiments lies in accurately measuring dynamics over mesoscopic length scales. Both methods leave a significant "blind spot" in the measurement of dynamics occurring over the tens to hundreds of nanometer range. To address this gap, Rheinstädter and coworkers[83] have proposed employing incoherent NSE spectroscopy to study lipid diffusion at mesoscopic length scales. NSE in general allows for probing the coherent dynamics of soft matter systems over long timescales (on the order of hundreds of nanoseconds). However, there are challenges associated with this approach, including the need to ensure that the incoherent spin-echo signal is large enough to be detected over the coherent scattering signal. In reconciliation, it's essential to acknowledge that macroscopic fluorescence and microscopic QENS methods observe different diffusion processes, which explain the variations in the measured diffusivity. However, even at short length scales up to a few nanometers, the precise mechanism of lateral diffusion remains highly contentious. Various models, such as jump diffusion[69, 84], ballistic flow-like motion[51, 67], continuous diffusion[47, 68], localized translational motion[54], hopping diffusion[85, 86], and sub-diffusive motions[65, 66], have been proposed to elucidate lateral lipid motion. Early QENS measurements suggested either continuous Brownian diffusion[87] or jump diffusion[84] of lipids within the membrane. However, a new perspective on short-time molecular motion emerged when Falck et al. (2008) analyzed the motion of individual lipids with respect to their neighbours. It was observed that when an individual lipid molecule exhibited rapid movement in a particular direction, its neighbouring molecules also tended to move in the same direction, resulting in the formation of lipid clusters that drifted collectively for a limited time. These clusters transiently formed, lasting only up to nanoseconds[67], before disintegrating and randomly reorganizing elsewhere. This motion is termed as a collective flow-like motion in which lipid molecules move synchronously with their neighbouring molecules, forming loosely bound clusters that moved together in a consistent direction for a brief period.

      This flow-like motion, depicted in Fig. 4, contrasts with continuous diffusion, which involves the Brownian motion of individual lipid molecules. A subsequent QENS experiment by Busch et al.[51] on hydrated lipid powders without solid support suggested that ballistic flow-like motion might be a more efficient search strategy than continuous Brownian motion. Bayesian data analysis has suggested that the fundamental mechanism underlying long-range diffusion in lipid membranes is characterized by flow-like motion. However, Armstrong et al.[86] found a different mechanism for lateral diffusion in single lipid bilayers supported by solid substrates. Their study suggested continuous diffusion of lipid rather than the flow-like ballistic motion reported in stacked membrane systems. They also observed hopping diffusion, an enhanced diffusion at the nearest neighbour distance of the lipid molecules, likely attributed to the effect of the supporting substrate on lipid organization. The study of lipid diffusion on very short length scales requires access to high $Q$ values. For this purpose, the thermal backscattering spectrometer IN13 at the Institut Laue-Langevin (ILL) was used, which is capable of measuring $Q$ values from 0.2 to 4.9 Å$^{-1}$ (~1.3 - 31.5 Å). At low $Q$ values, observed quasielastic broadening is described by a Lorentzian. The width of the Lorentzian increases linearly with $Q^2$ suggesting continuous diffusion. However, at higher $Q$ values > 2.5 Å$^{-1}$, quasielastic broadening is better described by a Gaussian instead of Lorentzian and the width of Gaussian increases linearly with $Q$. This revealed a change in dynamics at higher $Q$ values > 2.5 Å$^{-1}$ or a length scale smaller than 2.4 Å (less than the lipid diameter), indicating a shift from continuous diffusion to ballistic like motion. Typical fitted QENS data at higher $Q$ values namely 2.65 Å$^{-1}$ and 3.59 Å$^{-1}$ are shown in Fig. 6 (a) and (b), respectively. Fig. 6 (c) shows the FWHM for all measured $Q$ values as a function of $Q^2$. It is evident that at low $Q$ values < 2.5 Å$^{-1}$ (blue circle), FWHM is linear with $Q^2$. Then there is a sudden decrease in the quasielastic broadening. This broadening is described using a Gaussian, width of

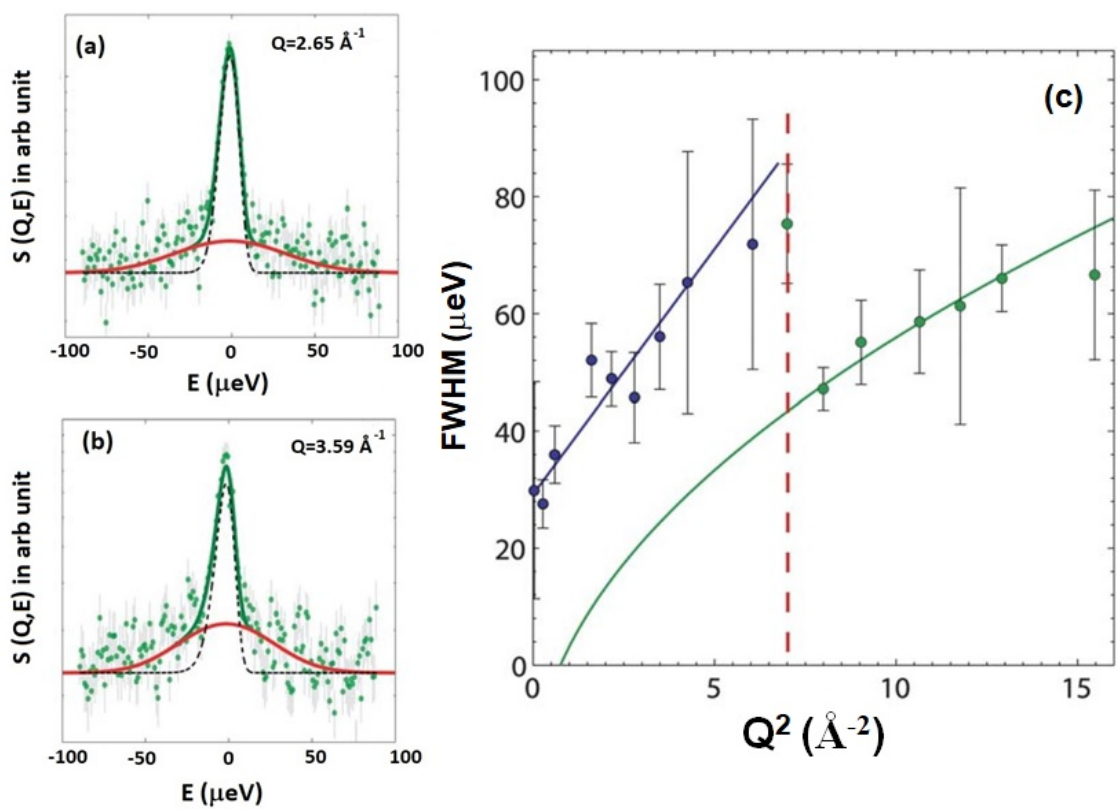

Fig. 6 Typical fitted QENS data with an elastic line (black line) and a Gaussian quasielastic broadening (red line) at (a) $Q = 2.65$ Å$^{-1}$ and (b) $Q = 3.59$ Å$^{-1}$. (c) Full Width at Half Maximum (FWHM) of the quasielastic broadening plotted as a function of $Q^2$. At low $Q$ values (blue circles), the quasielastic broadening is described by a Lorentzian with FWHM linearly dependent on $Q^2$, suggesting continuous diffusion. The corresponding fit is shown by blue line. The high $Q$ data (green circles) exhibit quasielastic broadening described by a Gaussian, with FWHM increasing linearly with $Q$. The corresponding fit (green) displays a square root relationship when plotted against $Q^2$. The transition between continuous and ballistic diffusion is marked by the red dashed line. (adapted from ref.[68]).

which increases linearly with $Q$. From this linear dependence, a lipid velocity of 1.1 m/s was extracted, which is notably two orders of magnitude lower than the purely ballistic velocity (87 m/s) of a free particle with the mass of the lipid molecule. This discrepancy suggests that the short-range lipid motion exhibits characteristics more akin to flow-like behavior rather than purely ballistic motion. Consequently, this prompts consideration of the potential influence of "nano-viscosity" or "nano-friction" on lipid dynamics at length scales smaller than the nearest-neighbour distance. In summary, at the nanometer length scale, lipid molecules exhibit continuous diffusion. However, when examining at length scales shorter than approximately one-third of the lipid nearest neighbor distance, there is a discernible change in the character of motion. At these shorter ranges, lipid motion displays flow-like attributes instead of adhering to continuous diffusion or a rattling-in-the-cage motion. These findings align with QENS results on unilamellar vesicles (ULVs) by Sharma et al.[22], which observed continuous Brownian diffusion of lipids over the nanometer length scale. In contrast, QENS findings on hydrated supported bilayers by Wanderling et al.[54] revealed that lateral lipid diffusion is characterized as localized translational diffusion within a confined cylindrical region.

Comparatively, MD simulation studies[37, 65, 66] have shown that the MSD corresponding to the lateral motion of lipids exhibits a power-law dependence of $t^\alpha$ ($\alpha < 1$), suggesting anomalous lateral diffusion. For example, MSD for DODAB lipid as obtained from MD simulation in both ordered and fluid phases are shown in Fig. 7 (a). MSD was found to follow a sub diffusive regime with a dependence of $t^\alpha$ in both the ordered and fluid phases. Here $\alpha$ is sub-diffusive exponent which can be explicitly calculated by the logarithmic time derivative of MSD. Variation of $\alpha$ with time is shown in Fig. 7(b). It is evident that $\alpha < 1$ (at long time) for both the ordered and fluid phases suggesting lateral motion of lipid is subdiffusive. The sub-diffusive motion of lipids in the membrane, associated with crowding of lipids, is described as a non-Markovian diffusion process. The generalized Langevin equation (GLE) with a power-law memory kernel is employed to model the non-Markovian diffusion behavior of lipid molecules[66]. This approach is particularly suitable for representing the lateral diffusion of lipid. The power-law memory kernel provides a robust framework for capturing the observed sub-diffusive characteristics of lipid movement, as indicated by MD simulations. By incorporating these findings, the GLE-based model has been refined to accurately interpret QENS data as shown in Fig. 7 (c), offering deeper insights into the dynamics of lipid diffusion within membranes. Variation of relaxation rate corresponding to lateral diffusion, $\Gamma_{lat}$ with $Q$, is shown in Fig. 7 (d). For direct comparison, $\Gamma_{lat}$ obtained from MD simulation is also shown in the figure. It is evident that both are consistent with each others. The lines indicate fitting based on the quadratic dependence $1/4AQ^2$. The value of $A$ obtained from the least-squares fit was found to be 0.42 Å$^2$/ps$^\alpha$ ($\alpha$=0.61) and 0.34 Å$^2$/ps$^\alpha$ ($\alpha$=0.61) for MD simulation and QENS experiments, respectively.

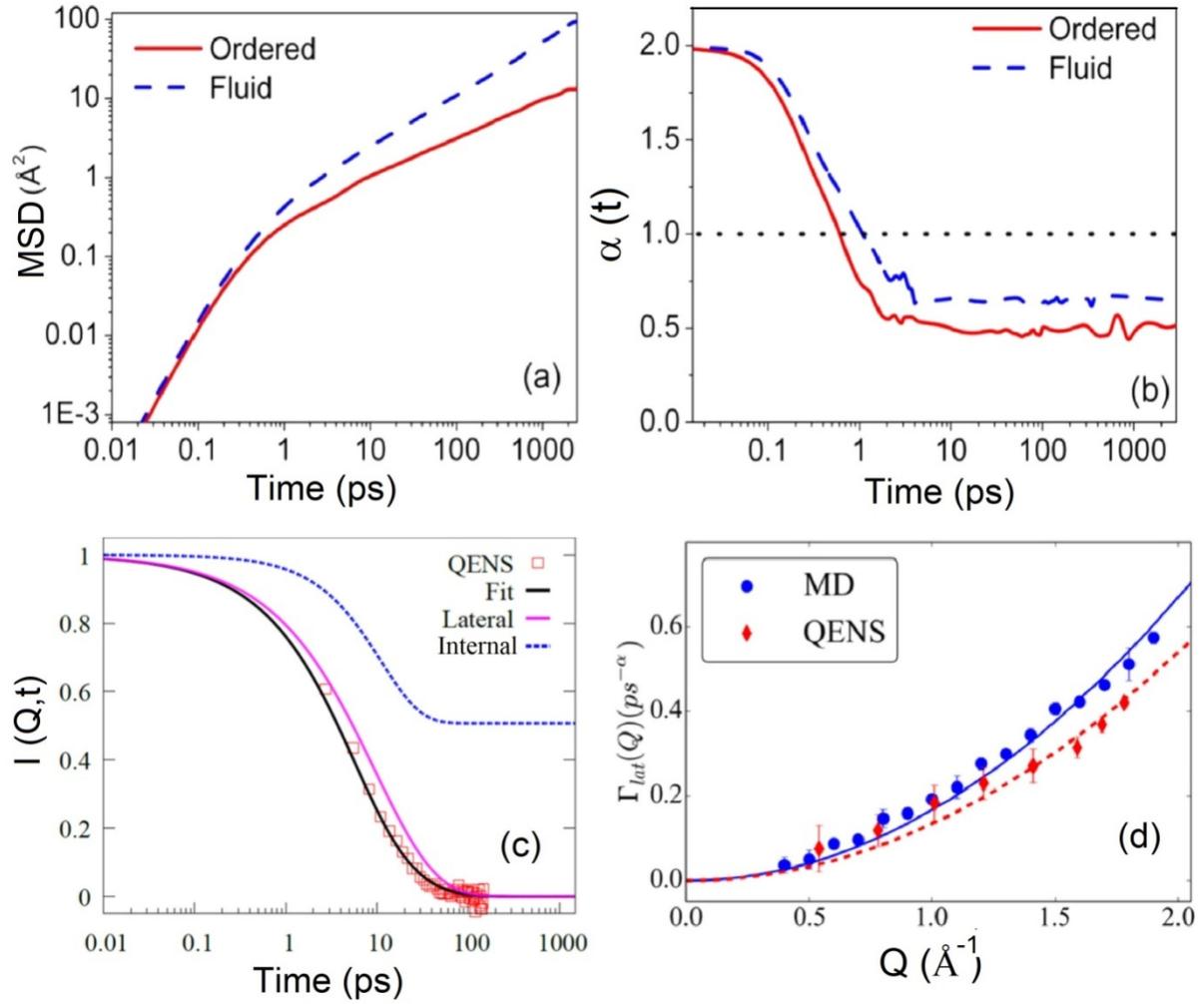

Fig. 7 (a) Mean square displacement (MSD) for lateral motion of DODAB lipid in the ordered (300 K) and fluid (350 K) phases. (b) Variation of the sub-diffusive exponent (α) with time in both ordered and fluid phases. (c) Incoherent intermediate scattering function obtained from Fourier transform of QENS spectra at $Q = 1.2$ Å$^{-1}$ in the fluid phase (345 K). Individual components corresponding to lateral and internal motion are also displayed alongside the fits. (d) Variation of the decay constant ($\Gamma_{lat}$) corresponding to lateral motion in the fluid phase, obtained from the fits shown in Fig. 7 (c). For direct comparison, $\Gamma_{lat}$ obtained from MD simulation is also depicted (adapted from ref.[66]).

It is evident that the process of lateral diffusion is intricate and not yet fully understood. A meaningful comparison of lateral diffusion in membranes requires consistent measurement techniques, as the diffusion coefficient is highly sensitive to both the observation time and the length scales involved. Generally, macroscopic techniques yield diffusion coefficients that are an order of magnitude slower compared to those obtained from microscopic techniques. Even within the same method, such as microscopic QENS, care must be taken when comparing lateral diffusion coefficients from different instruments. Each QENS instrument has unique instrumental resolution characteristics, hence corresponding distinctive observation time scales, which can significantly affect the measured diffusion values. It has been shown that lower instrumental resolution and longer observation times tend to give slower diffusion coefficients, indicating the importance of understanding the specific measurement context[16, 88]. Additionally, several factors related to the lipid bilayer (such as area per lipid, molecular structure and charge of lipid, membrane viscosity), solvent viscosity and temperature can influence the lateral diffusion. These factors can either facilitate or hinder the movement of lipids across the membrane, thereby affecting cell membrane dynamics and functions. In the next section, we will examine in details how these factors impact the lateral diffusion of lipids.

**FACTORS AFFECTING THE LATERAL DIFFUSION**

**(i) Area per lipid:**

The area per lipid is an important parameter of any lipid bilayer, representing the average cross-sectional area occupied by each lipid molecule within the bilayer. It can be calculated by dividing the total cross-sectional area of the bilayer, measured along the XY plane (the plane parallel to the bilayer surface), by half the total number of lipid molecules in the bilayer. An increase in area per lipid generally promotes faster lateral diffusion as there is more space for individual lipids to move within the membrane. Area per lipid increases with the increase in temperature[89]. In particular, a sharp jump in the area per lipid is observed at the main phase transition of the lipid membrane. For example, in the case of DMPC, the area per a lipid molecule increases about 28 % from 47 $Å^2$ to 60 $Å^2$, across the gel to fluid phase transition[89, 90]. Sharma et al[47] examined the impact of gel-to-fluid phase transitions on the lateral diffusion of lipids in ULVs. Typical observed QENS spectra for DMPC membrane in the gel (280 K) and fluid (310K) at $Q=1.1$ $Å^{-1}$ are shown in Fig. 8 (a). For direct comparison, instrument resolution as observed using vanadium is also shown. Significant quasielastic broadening is evident for DMPC membrane in both gel and fluid phases suggesting presence of stochastic dynamics of lipids in membrane. A sharp increase in quasielastic broadening is found when membrane goes from gel to fluid phase. The QENS spectra can be effectively modeled assuming long range lateral diffusion and localized internal motion of lipids in both gel and fluid phases. Typical fitted QENS spectra for DMPC membrane in the fluid phase along with the individual components for lateral and internal are shown in Fig. 8 (b). Obtained HWHM of Lorentzian functions associated with lateral motions for both gel (280

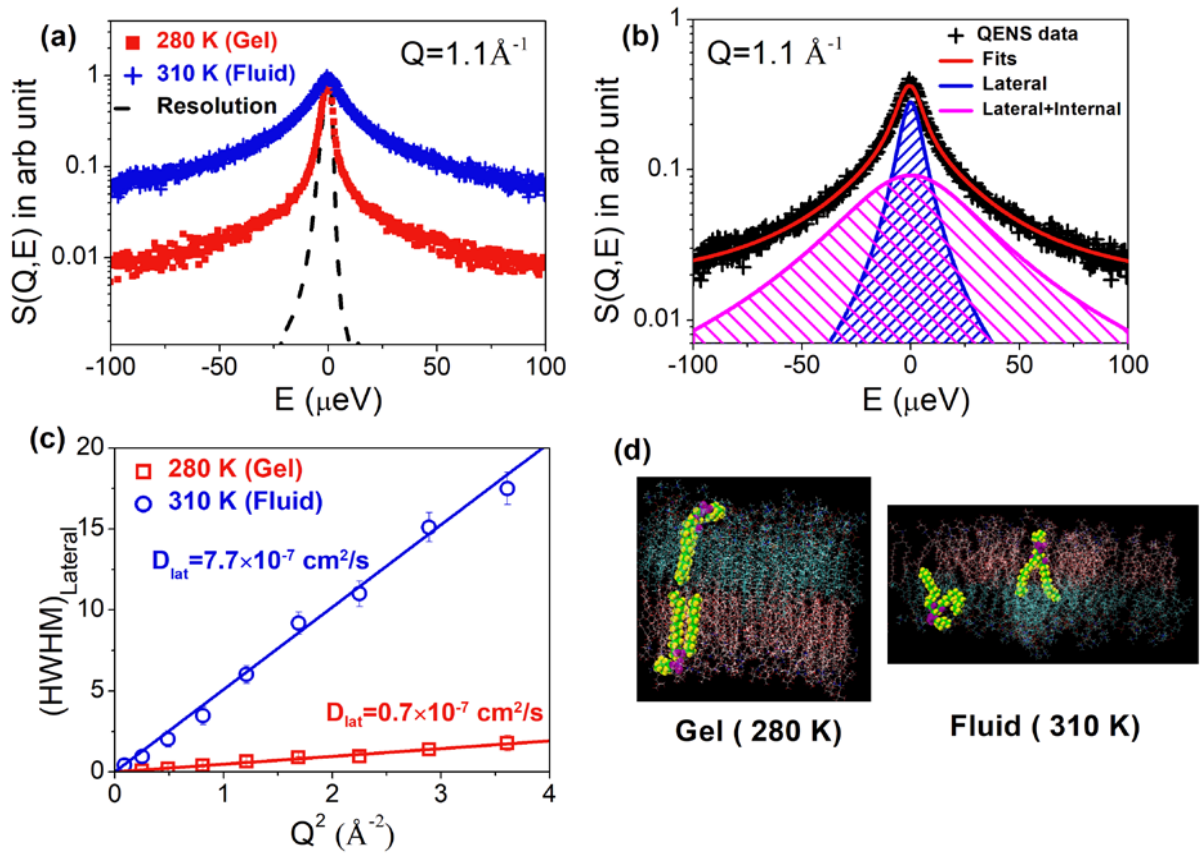

Fig. 8 (a) Typical observed QENS spectra for DMPC membrane in the gel (280 K) and fluid (310 K) phase. Instrument resolution as measured using vanadium standard is also shown. Spectra are normalized to peak amplitudes for the direct comparison. (b) Fitted QENS spectra for DMPC membrane at 310 K (fluid phase) using a scattering law assuming both lateral and internal motions of the lipid. Individual components corresponding to the lateral (narrower) and internal (broader) motions are also shown. (c) Variation of half width at half maxima (HWHM) of Lorentzian function associated with lateral motion with $Q^2$ at 280 K (gel) and 310 K (fluid) phase. It Is evident that in both the phases, HWHM follow continuous Ficks diffusion ( HWHM=$D_{lat}Q^2$). (d) MD simulation snapshot of DMPC at 280 K (gel) and 310 K (fluid). It is evident that in gel phase, lipid molecules are highly ordered with low area per lipid whereas in the fluid phase, they are highly disordered with high area per lipid. (Adapted from ref.[47]).

K) and fluid (310 K) phases are shown in Fig. 8 (c). It is evident that in both the phases, HWHM follows continuous Fick's diffusion model (HWHM=$D_{lat}Q^2$), withthe HWHM for the fluid phase being much higher than that of the gel phase. This is reflected in the respective lateral diffusion coefficients, as in the fluid phase (310 K), $D_{lat}$, was determined to be $(7.7 \pm 0.3) \times 10^{-7}$ cm$^2$/s, which was an order of magnitude higher than the value measured in the gel phase ( 280 K), i.e., $(0.7 \pm 0.1) \times 10^{-7}$ cm$^2$/s. These observations clearly vindicate that the main phase transition is closely linked to the microscopic lateral motion of lipid molecules. Aforementioned, at main phase transition, there is a sharp jump in area per lipid which promotes enhanced lateral diffusion, indicating a close relationship between the mobility of phospholipid molecules and the available area per lipid molecule. MD simulation snapshot of DMPC membrane in the gel (280 K) and fluid (310 K) phases are shown in Fig. 8 (d). In the gel phase, lipids are organized into a densely packed and orderly array, limiting their conformational, rotational, and translational freedom. However, during the main phase transition, the alkyl chains become disordered, exhibiting numerous gauche defects. This disorder leads to a significant increase in the area per lipid molecule, reducing molecular packing density. Notably, no significant changes in lateral diffusion coefficients were observed during the pre-transition from the gel to ripple phase. Even in the fluid phase, an increase in temperature results in a consistent rise in the area per lipid. For instance, in the fluid phase of DMPC, at 30°C, the area per lipid measures 60 Å$^2$, increasing to 66 Å$^2$ at 60°C. Notably, the lateral diffusion coefficient of lipid increases monotonously with temperature in the fluid phase[91].

**(ii) Hydrations**

Biological systems predominantly operate in aqueous environments, highlighting the significant influence of hydration levels on their dynamics. Hydration can be precisely regulated in case of supported lipid multilayers hence to investigate effects of hydration most of the experiments have been carried out on the supported multilayer. The main transition temperature and bilayer thickness of model membrane systems tend to decrease with increasing hydration[92, 93]. This hydration-dependent variation impacts lipid packing density, subsequently modulating the dynamics of lipids in multilayers. König et al.[92] explored the hydration-dependent dynamics of DPPC multilayers through QENS, covering hydration levels from 8 to 20 wt. %, corresponding to 3-10 water molecules per lipid molecule. In the fluid phase, lipid dynamics involve lipid diffusion within its solvation cage and kink diffusion, whereby the solvation cage size expanding as hydration increases. Lateral diffusion has been observed to rise with increasing hydration[62]. NMR experiments demonstrated a two-fold increase in DPPC diffusion in the liquid phase when hydration levels were elevated from 15% to 40% (*w/w*). Theoretical calculations suggest that hydration-dependent lipid diffusion largely arises from alterations in bulk water mobility within the multilamellar phospholipid/water system. Trapp et al.[93] emphasized hydration-dependent dynamics of head group motion by utilizing tail-deuterated lipids, enabling focused study on the head group. They observed a pronounced influence of hydration on head group mobility.

**(iii) Structure of membrane system**

The lateral diffusion of lipids has been extensively studied across various model membrane systems, including multilamellar vesicles (MLVs)[41, 94, 95], unilamellar vesicles (ULVs)[20, 27, 38, 42, 43, 47, 49, 50, 55, 56, 78, 79, 91, 96-100], supported lipid bilayers[54, 68, 84, 87], unsupported multilamellar stacks (hydrated powder)[51, 96], and anhydrous lipid powder[101]. Dynamics of lipid often vary significantly depending on the structure of the model membrane system, even when composed of the same lipids. For instance, in supported single bilayers, the lipid leaflet in direct contact with the substrate can be influenced by the solid support. This influence may lead to artifact results, such as modulated diffusion at the nearest neighbor distance of lipid molecules or suppression of the main phase transition[39]. Busch et al.[96] found that the lateral lipid mobility increases from multilayers to single bilayers and then to monolayers. They utilized three different model systems: multibilayers, single bilayers in vesicles, and monolayers in emulsions. Their results showed a 1.3-fold increase in mobility from multibilayers (hydrated powder) to single bilayers (ULVs) and a 2-fold increase from multibilayers to monolayers. Furthermore, they noted that the mobility of lipids in single bilayers (ULVs) increases as the diameter of the ULV decreases. The use of various membrane structures provides distinct advantages for understanding cell membranes. For example, supported membranes allow for the separation of motion within the membrane plane from motion outside of it. This separation can be achieved through in-plane and out-of-plane scattering geometries, assuming the lipid orientation in the supported membrane samples remains consistent over dimensions commensurate with the incident neutron beam (typically a few centimeters). Additional benefits of supported membranes include their stable surface and smooth morphology, which enable atomic force microscopy to reveal topological features at an atomic scale, such as nanometer-scale domains. These systems hold promise for various applications, including biosensors. On the other hand, studying lipids that self-assemble into vesicles in aqueous environments offers the advantage of closely mimicking the lipid arrangement found in cell membranes. This morphology provides valuable insights into the behavior and properties of lipid membranes in biological contexts.

**(iv) Membrane curvature:** Membrane curvature can create local variations in lipid packing and affect lateral diffusion. Regions with higher curvature have slower diffusion compared to flatter regions. Theoretical estimates of the diffusion coefficient of molecules on a curved surface, mimicking the microvilli observed in plasma membranes, revealed a slower diffusion as compared with an ideal planar surface[102].

**(v) Molecular Structure of lipid: Size and charge of polar head group, length and degree of unsaturation of alkyl tails**

The size and charge of lipid headgroups can affect lateral diffusion by influencing lipid-lipid interactions and packing. For ex. DMPG and DMPC has the same alkyl tails and differs only in the head group. DMPC has zwitterionic PC head group whereas DMPG has anionic PG head group. Both the lipids have the similar main phase transition temperature. Sharma et al., have shown that in the fluid phase, lateral diffusion coefficient for the zwitterionic DMPC membrane is much higher than that for the charged DMPG membrane[49]. Hence, one can infer that higher surface charge density of the membrane or Coulombic interactions between the head groups restricts the lateral diffusion of the lipids. This is further supported by the fact that addition of NaCl which reduces surface charge density and screens the Coulumbic repulsion between the head groups enhances the lateral diffusion[49]. Increasing the degree of unsaturation in the acyl chains of the lipids is generally observed to enhance membrane fluidity or lateral diffusion. However, comparing the lateral diffusion coefficients of lipids with varying alkyl chain lengths or degrees of unsaturation poses a significant challenge due to their different melting temperatures ($T_m$) as it is highly sensitive to the acyl chains. $T_m$ tends to increase with longer acyl chain lengths and decrease with higher unsaturation levels in the lipid's acyl chain. To address this challenge, researchers often use a reduced temperature parameter, $T_r=(T-T_m)/T_m$. Studies have shown an inverse relationship between acyl chain length and lateral diffusion for saturated PC, with the fastest lateral diffusion observed for DSPC and the slowest for DLPC at the same reduced temperature[72].

**(vi) Solvent viscosity and ionic strength**

In the extended free volume theory and continuum hydrodynamic model, it is evident that solvent viscosity also plays an important role in the lateral diffusion of the lipid. Lateral diffusion coefficient is inversely proportional to the solvent viscosity. To examine this, Sharma et al., have carried out QENS experiments on DMPC membrane in two different solvents namely (i) pure $D_2O$ and (ii) a salt solution of $(LiCl)_{0.13}(D_2O)_{0.87}$ or 7.6 m LiCl[79]. Using QENS, we found that at 310 K, the lateral diffusion of DMPC in $(LiCl)_{0.13}(D_2O)_{0.87}$ is five times slower than in pure $D_2O$. Our QENS measurements indicate that water translation diffusivity in $D_2O$ at 310 K is 2.4 times faster than in $(LiCl)_{0.13}(D_2O)_{0.87}$, suggesting higher viscosity in the latter. Increased solvent viscosity, coupled with salt effects from monovalent cation interaction with lipid carbonyl oxygen atoms, mainly accounts for reduced lipid lateral diffusion in $(LiCl)_{0.13}(D_2O)_{0.87}$. Furthermore, the local diffusivity of lipid tails decreased by a factor of 2.2 in the presence of salt, correlating with increased water viscosity in the LiCl solution. The concentration of ions in the surrounding environment can influence the strength of electrostatic interactions, impacting the diffusion behavior of charged lipids. For example, Sharma et al., examined anionic DMPG vesicles in $D_2O$ and 100 mM NaCl $D_2O$[49]. DMPG vesicles in $D_2O$ exhibit an unusually broad melting range between the gel and fluid phases, characterized by high viscosity due to vesicle size and ionization increases. In this anomalous regime, vesicle viscosity is approximately ten times higher than in the ordered and fluid phases. However, our QENS results show no abrupt changes in membrane dynamics, suggesting that macroscopic viscosity measurements do not directly correlate with nanoscale membrane dynamics[49]. While the DMPC study [79] emphasized the role of solvent viscosity in modulating membrane dynamics, in the case of DMPG, a significant difference in macroscopic vesicle solution viscosity did not markedly affect lateral diffusion. It's worth noting that despite differences in macroscopic vesicle solution viscosity, the viscosities of both $D_2O$ and 100 mM NaCl $D_2O$ solvents were similar. Our findings suggest that nanoscale lipid lateral motion is largely independent of macroscopic vesicle solution viscosity but is influenced by the solvent viscosity.

**(vii) Effects of temperature and pressure**

As with any thermally activated motion, lateral diffusion is accelerated by increasing temperature. Temperature also affects other membrane properties, such as the area per lipid and solvent viscosity, both of which influence lateral diffusion. Thus, when considering all these factors, the lateral diffusion coefficient generally increases with rising temperature. Additionally, if a temperature change leads to a shift in the membrane's physical state—for instance, from an ordered phase to a fluid phase—there can be a substantial increase in lateral diffusion, often by an order of magnitude.

The impact of hydrostatic pressure on the dynamics of DMPC MLVs has been studied, revealing distinct effects depending on the membrane phase[94]. In the fluid phase, increased pressure significantly impedes lipid mobility, indicating that pressure can affect the fluidity and motion of lipids within the membrane. In contrast, when the membrane is in the gel phase, the dynamics show relatively little change with increasing pressure, suggesting that lipids are already constrained in this more rigid state. These findings underscore the role of pressure in modulating membrane dynamics and provide valuable insights into the stability and behavior of lipid membranes under varying conditions.

**(ix) Membrane compositions**

It's well-established that alterations in membrane composition can exert significant effects on lipid lateral diffusion. For instance, the inclusion of cholesterol has been demonstrated to diminish the available free area for lipids, resulting in a reduction in lateral diffusion[103]. Moreover, the impact of cholesterol on lipid lateral diffusion appears to be concentration-dependent, with higher concentrations further impeding the mobility of lipids within the membrane. Beyond cholesterol, other components of the membrane, such as specific lipid species or membrane proteins can introduce steric hindrance or alter lipid-protein interactions, thereby impacting lateral diffusion behavior. Changes in the composition of the membrane also modify the interaction between its constituents, thereby modulating the activation energy and consequently affecting the lateral diffusion of lipids.

In this section, we've explored various factors that influence the lateral diffusion of lipids. All these factors are summarized in the Table-1. In the next section, we'll delve into how the presence of membrane-active compounds impacts lipid lateral diffusion.

**Table: 1 Effects of physical factors on the lateral diffusion of the lipids**

| S. No | Factor | Effects on $D_{lat}$ | Refs. |
|---|---|---|---|
| (i) | Area per lipid (APL) | • $D_{lat}$ increases with the APL | 47 |
| (ii) | Hydration | • $D_{lat}$ increases with hydration | 62, 93 |
| (iii) | Structure of membrane system | • Increase in mobility from multibilayers to single bilayers by a factor of 1.3, and to monolayers by a factor of 2<br><br>• Mobility of the lipid in single bilayer (ULVs) increases with decreasing the diameter of the ULVs<br><br>• Supported vs unsupported membrane system; direct contact with the substrate modulate the lateral diffusion | 96 |
| (iv) | Membrane Curvature | • $D_{lat}$ get slower with increase in the membrane curvature. Hence, highly curved has slower diffusion compared to flatter regions. | 102 |
| (v) | Molecular structure of the lipid | • Size and charge on head group affects the lateral diffusion. For Ex. charged vs zwitterionic, in case of charged lipid, lateral diffusion will be slower than zwitterionic<br><br>• Increase in alkyl chain length in the fluid phase, no significant effects on the lateral diffusion<br><br>• Consistent with free volume model | 49<br><br>61 |
| (vi) | Solvent viscosity and ionic strength | • As per extended free volume model, $D_{lat}$ should decrease with an increase in the solvent viscosity<br><br>• Addition of salt affects the lateral diffusion | 79<br><br>49 |
| (vii) | Temperature and Pressure | • Increase in temperature leads to increase in $D_{lat}$<br><br>• $D_{lat}$ increases by an order of magnitude faster when membrane physical state changes from ordered phase to fluid phase<br><br>• Increase in pressure restricts dynamics significantly in the fluid phase<br><br>• Relatively small change in $D_{lat}$ with pressure when membrane is in the gel phase | 47<br><br>94 |
| (viii) | Obstruction effects | • The significant reduction in the diffusion coefficient due to obstacles | 63 |
| (ix) | Membrane composition | • Change in membrane composition affect free area and $E_a$ that affect $D_{lat}$ | |

## EFFECTS OF MEMBRANE ACTIVE COMPOUNDS ON THE LATERAL DIFFUSION

Cell membranes serve as dynamic platforms for signaling, transport, and molecular recognition. However, this dynamic equilibrium can be perturbed by the presence of membrane-active compounds, ranging from therapeutic drugs to stimulant to environmental toxins. Membrane active compounds can impact various membrane properties such as thickness, curvature, main phase transition temperature, and area per lipid, all of which directly influence lipid lateral diffusion, as discussed earlier. Additionally, this compound can alter interactions between the constituents of the membrane, further affecting lateral diffusion. Studies indicate that long-range repulsive potentials decrease $D_{lat}$ by effectively expanding particle radii, whereas attractive potentials reduce $D_{lat}$ by promoting particle clustering or adhesion to obstacles[104]. Any alterations in membrane dynamics induced by external compounds can consequently impact these vital cellular functions. Understanding the intricate interplay between these compounds and cellular membranes is paramount, not only for unraveling the mechanisms of their action but also for advancing our knowledge of membrane biophysics and its implications for health and disease. Membrane dynamics also influence the behavior of synthetic membrane systems used in drug delivery, diagnostics, and biomimetic applications, highlighting the importance of understanding how membrane-active compounds affect their functionality and performance. Therefore, investigating the effects of membrane-active compounds on membrane dynamics holds significant implications for both basic research and practical applications in biomedicine and biotechnology. In this section, we will summarize interaction of various membrane active compounds including peptides or proteins, drugs, antioxidants, stimulants and antimicrobials.

**(i) Membrane Active Peptides**

Membrane-active peptides constitute a diverse group of peptides that dynamically interact with cellular membranes, eliciting effects such as disruption or penetration. These peptides are integral to various biological processes, including antimicrobial defense, cellular signaling, and drug delivery. Examples of membrane-active peptides encompass antimicrobial peptides, amyloid peptides, and cell-penetrating peptides. The interaction between these peptides and membranes can be examined by assessing their perturbing effects on the structural, dynamic, and phase behavior of the lipid membrane. In this section, our primary focus will be on elucidating the impact of these peptides on the lateral diffusion of lipids.

**(a) Amyloid Peptides**

Amyloid peptides represent a distinct class of proteins recognized for their tendency to assemble into stable β-sheet-rich structures, giving rise to formation of amyloid fibers which are integral to the pathogenesis of both neurodegenerative and metabolic disorders. Prominent examples include amyloid beta (Aβ), alpha-synuclein, and islet amyloid polypeptide (IAPP), associated with Alzheimer's disease, Parkinson's disease, and type II diabetes, respectively. Aβ peptides aggregate to create insoluble extracellular plaques in proximity to neuronal cells in the human brain, thought to play a crucial role in the development of Alzheimer's disease. Typically comprising 39–42 amino acids, Aβ peptides exhibit membrane-active properties. Buchsteiner et al. conducted a study on the effects of Aβ (25-35) on lipid dynamics in supported membranes composed of DMPC:DMPS (92:8)[105]. Their QENS measurements, performed in both the ordered and fluid phases, revealed that this neurotoxic Aβ fragment enhances lateral diffusion of the lipid membrane in both phases. Moreover, an augmented local mobility of the lipids was noted, with a more pronounced effect observed in the "out-of-plane" direction. Extending this investigation, Barrett et al.[88] examined the impact of another Aβ-derived peptide, Aβ (22-40), which has a different amino acid composition and length. Their QENS measurements, carried out on supported DMPC:DMPS (92:8) lipid membranes, revealed intriguing findings. While Aβ (22-40) accelerated lateral diffusion of lipids in the ordered phase, a contrasting effect was observed near the main phase transition temperature, where lipid diffusion decreased by 24%. Another study by Buchsteiner et al.[106] have observed an increase in lateral diffusion in the fluid phase. Our study[43] focused on the effects of Aβ (1-40) on the dynamics of anionic DMPG in ULVs. We found that in the fluid phase, Aβ (1-40) accelerates lateral diffusion but does not significantly affect localized internal lipid motion. The addition of Aβ (1-40) in the fluid phase led to effective thinning of the bilayer, as observed from small angle neutron scattering (SANS), potentially increasing the area per lipid and hence lateral diffusion coefficient. Interestingly, no significant effect on lateral diffusion was observed in the ordered phase. In summary, investigations into various Aβ-derived peptides, including Aβ (25-35), Aβ (22-40), and Aβ (1-40), using QENS, have shown that these peptides generally enhance lateral diffusion of lipid membranes in the fluid phase. However, near the main phase transition temperature, Aβ (22-40) exhibited a decrease in lipid diffusion, while Aβ (25-35) showed acceleration in the ordered phase. Conversely, Aβ (1-40) did not significantly impact lipid membrane dynamics in the ordered phase. These findings suggest that the interaction between Aβ peptides and lipid membranes is complex and context-dependent, with potential implications for the pathogenesis of Alzheimer's disease. Further research into the mechanisms underlying these interactions could provide valuable insights into the role of lipid membranes in the development and progression of Alzheimer's disease, potentially paving the way for novel therapeutic approaches targeting membrane-associated pathways.

**Antimicrobial peptides**

Antimicrobial peptides (AMPs) are emerging as promising candidates for combating multidrug-resistant bacteria due to their multifaceted mechanism of action[107]. AMPs exist in all living organisms, including plants, animals, and mammals. Some examples of AMPs are Melittin, alamethicin, and aurein, which are found in bee venom, the fungus Trichoderma Viride, and the Austrian frog, respectively. Cartoon representations of these AMPs are shown in Figure 9 (a). Melittin is a cationic peptide (+5e) consisting of 26 amino acids (AA). Alamethicin and aurein are anionic with 20 AA and cationic with 13 AA, respectively. Sharma et al.[48, 98] have shown that the inclusion of a minimal quantity of a prototypical cationic AMP, melittin, has a profound impact on the dynamical and phase behavior of membranes. Even at concentrations lower than those necessary for pore formation, melittin induces significant alterations in membrane dynamics, eliminating both pre-transition (gel-ripple) and main (ripple-fluid) transitions of the lipid membrane[48, 98]. Neutron elastic intensity scan for DMPC membrane with and without 0.5 mol % melittin are shown in Fig. 9 (b). The impact on membrane dynamics is profoundly influenced by the bilayer phase: in the ordered phase, melittin acts as a plasticizer, enhancing membrane dynamics, while in the fluid phase, it acts as a stiffening agent, constraining membrane dynamics[48, 98]. Obtained $D_{lat}$ for DMPC membrane in absence and presence of 0.5 mol % melittin at different temperature are shown in Fig. 9 (c). For direct comparison, $D_{lat}$ obtained for DMPC membrane in presence of 0.5 mol % alamethcin at different temperature are also shown in Fig. 9 (c). In the ordered phase, it is evident that alamethcin does not significantly affect lateral motion, whereas melittin enhances it. However, contrasting these results, in the fluid phase, both AMPs slow down lateral diffusion. Melittin acts as a stronger stiffening agent, more efficiently restricting lateral diffusion compared to alamethicin. These studies suggest that melittin's disruptive effects on lipid lateral motion, even at sub-pore-forming concentrations, could jeopardize cell stability by rendering the membrane more susceptible to additional stresses and defects[98]. This novel mechanism of action highlights how reduced lateral mobility induced by melittin can lead to changes in membrane properties, impeding membrane-related biological processes and ultimately contributing to bacterial cell death without pore formation or membrane destruction. Selectivity is a crucial characteristic of AMPs, ensuring their efficacy as therapeutic agents while minimizing toxicity to the host. One key distinction between bacterial and mammalian membranes lies in their composition, such as the presence of cholesterol in mammalian membranes but not in bacterial membranes. Sharma et al.,[48] have shown that the interaction of melittin with the phospholipid membrane strongly relies on the cholesterol content. In the presence of cholesterol, the destabilizing effect of melittin is mitigated [48]. Our research demonstrates that these AMPs selectively target bacterial membranes without disrupting mammalian membranes[48, 91], a crucial prerequisite for their antibiotic function. Another significant compositional difference is the prevalence of anionic phospholipids, such as PG and cardiolipin, in the bacterial membranes. In contrast, the outer leaflet of mammalian membranes predominantly consists of zwitterionic phospholipids like PC and SM. We investigated the effects of two ubiquicidin (UBI)-derived peptides, UBI(29-41) and UBI(31-38), which possess charges of +6e and +4e, respectively. Both peptides were found to restrict lipid lateral diffusion in the fluid phase of the membrane, with UBI(29-41)

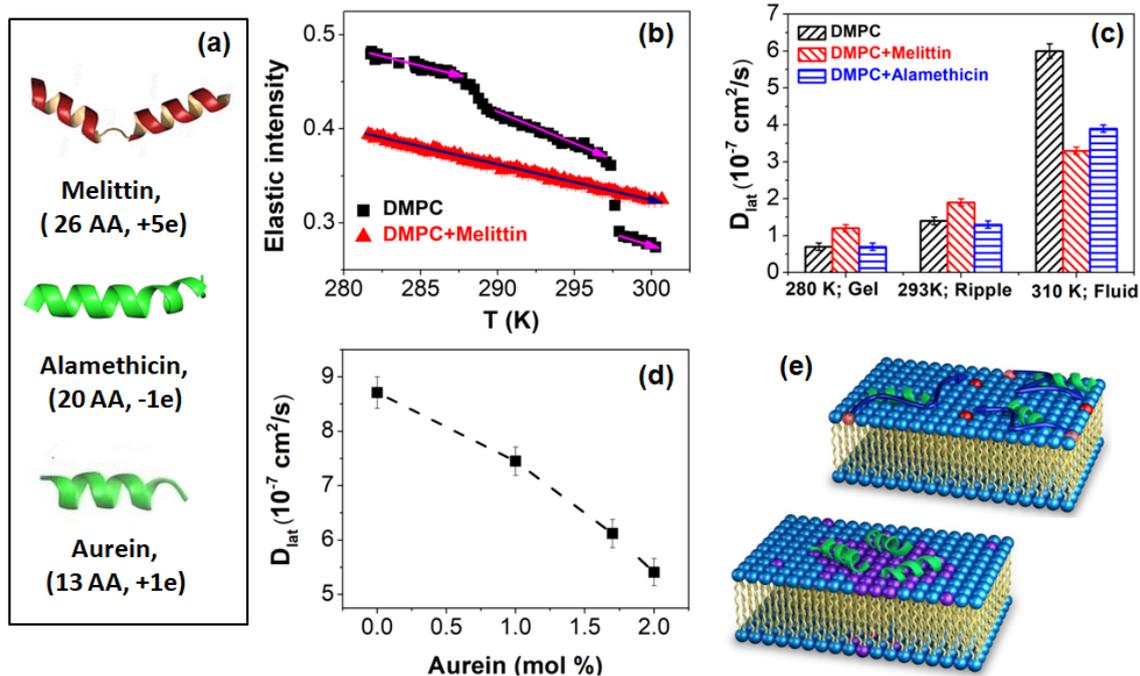

Fig. 9 (a) Cartoon representations of different antimicrobial peptides namely melittin ( 26 amino acid (AA)), alamethicin (20 AA) and aurein (13 AA) (b) $Q$-averaged elastic intensity scans for DMPC vesicles in absence and presence of 0.5 mol % melittin in heating cycle. (c) Lateral diffusion coefficient ($D_{lat}$) of DMPC lipid in absence and presence of 0.5 mol % melittin and alamethicin at different temperatures. (d) Variation of $D_{lat}$ of DMPC with aurein concentrations at 320 K. (e) Schematics of lateral diffusion and segregation modulated by AMP (adapted from refs. [91, 98]).

exhibiting more significant effects, suggesting a stronger interaction with the lipid membrane as confirmed by various biophysical measurements. Furthermore, our study revealed that ubiquicidin-derived peptides selectively bind to anionic phospholipid membranes, resembling bacterial membranes, and predominantly reside on the membrane surface[27, 108]. This property of the peptides holds potential for their application as in-situ infection imaging probes. Sharma et al.[98] have also investigated effects of other AMP such as aurein 1.2 on the membrane dynamics. While aurein had minimal impact on lateral diffusion in the ordered phase, akin to melittin, it restricted lateral diffusion in the fluid phase. Notably, lateral diffusion decreased consistently with increasing concentrations of aurein as shown in Fig. 9 (d). Comparative analysis among various AMPs revealed that melittin exhibited the strongest interaction with the membrane and the greatest restriction of lateral diffusion. Furthermore, melittin displayed exceptional behavior, acting as a plasticizer that enhanced membrane dynamics in the ordered phase. Considering cell membrane comprises various lipids, we created a homogeneous mixture of zwitterionic and anionic lipids. Employing contrast matching SANS technique, Sharma et al., investigated the effects of AMPs, such as aurein 1.2, on bilayer structure[91]. For the first time, their study has revealed the induction of lateral lipid segregation in the membrane by an AMP[91]. The formation of nanodomains can be attributed to reduced lateral lipid diffusion. The slower lateral diffusion, may lead to demixing of lipids or coexistence of phases. Enhanced interaction between charged lipids and peptides could exacerbate this effect, resulting in the observed nanoscopic lateral domains in the contrast matching SANS experiment. Schematics of modulated lateral diffusion and lateral segregation are shown in Fig. 9 (e).

2. **Transmembrane peptides**

Transmembrane proteins are integral membrane proteins that traverse the entirety of the cell membrane. The impact of a transmembrane sequence from the transferrin receptor (TFRC) protein on the dynamics of DMPC membranes has been investigated[44]. TFRC plays a crucial role in iron transport into the cell[109] and in regulating cellular iron balance. When coupled with apotransferrin, TFRC accumulates in specialized membrane regions and facilitates iron ion delivery via receptor-mediated endocytosis. To facilitate protein accumulation in specific membrane regions, proteins must exhibit lateral mobility within the membrane plane. To examine protein dynamics, vesicles containing chain-deuterated DMPC-d54 were used[44]. QENS experiments were conducted on two sets of samples: pure DMPC vesicles (DMPC and DMPCd54) and analogous samples with an additional 6 mol% of TFRC peptide, at 310 K using IN16B (ΔE=0.75 μeV) [44]. This allows for the experimental determination of the lateral self-diffusion coefficient of lipid molecules in the absence and presence of TFRC peptides, as well as the calculation of the self-diffusion coefficient of TFRC peptides. Data analysis utilized models corresponding to both continuous diffusion[47, 68] and flow-like motion[51, 67], with the diffusive motion model demonstrating superior fit, indicating that lipid long-range motion is best described by diffusion processes over observation times as short as $t_{obs} \approx 5$ ns (corresponding to ΔE=0.75 μeV). Deuteration of lipid chains was found to have no impact on the long-range dynamics of lipid molecules, with a lateral diffusion coefficient for DMPC measured at $2 \times 10^{-7}$ cm$^2$/s. However, the presence of TFRC restricted lateral diffusion, reducing the lateral diffusion coefficient to $0.8 \times 10^{-7}$ cm$^2$/s. At 6 mol% TFRC, peptide scattering contribution for chain-deuterated DMPC-d54 ULVs was approximately 50%, providing insight into peptide dynamics. The self-diffusion coefficient of TFRC peptides was determined to be $D_{peptide} = 0.5 \times 10^{-8}$ cm$^2$/s, approximately 40 times smaller than that of lipids. This finding supports the interpretation of reduced lipid long-range mobility in the presence of TFRC. Transmembrane peptides are hindered in their diffusion not only due to their larger mass compared to lipids but also because of their anchoring in both membrane leaflets. Consequently, neighboring lipid molecules are likely hindered in their long-range mobility, leading to a decrease in their self-diffusion coefficient.

## 3. Nonsteroidal anti-inflammatory drugs (NSAIDs)

NSAIDs are commonly prescribed for their antipyretic and pain-relieving properties, despite exhibiting a wide range of side effects. The notion that NSAID effects are associated to their interactions at the cellular membrane level has spurred investigations into their membrane interactions. Moreover, given the pivotal role of membrane dynamics in drug encapsulation and intracellular delivery, comprehending the dynamic and structural changes induced by NSAIDs in membranes is imperative for precise targeted delivery strategies. Sharma and co-workers have investigated the impacts of three widely used NSAIDs—aspirin, ibuprofen, and indomethacin—on the biophysical properties of model plasma membranes[56, 57, 97, 110, 111]. Chemical structures of these NSAIDs are shown in Fig. 10 (a). Differential scanning calorimetry (DSC) and neutron elastic intensity scan measurements revealed notable alterations in membrane phase behavior induced by these drugs. Typical observed neutron elastic intensity scans for DMPC membrane in absence and presence of 25 mol % NSAIDs are shown in Fig. 10 (b). It is evident that incorporation of NSAID led to shifts in the main phase transition towards lower temperatures and broadened transitions, indicating an influence on cooperativity[56, 97, 110, 112]. QENS data further unveiled enhanced membrane dynamics upon NSAID inclusion, particularly in the ordered phase, where all three NSAIDs enhanced lateral diffusion. Obtained $D_{lat}$ for DMPC membrane in absence and presence of 25 mol % NSAIDs at different temperatures are shown in Fig. 10 (c). It is evident that at 280 K and 293 K, inclusion of NSAID enhances $D_{lat}$. Notably, $D_{lat}$ increased monotonously with NSAID concentration. While the effects on membrane dynamics varied, quantitative differences were noted among the NSAIDs, correlated with their distinct interactions with lipid membranes[97, 110]. For instance, in the ordered phase, ibuprofen exhibited stronger effects (Fig. 10 (c)) on the membrane compared to other NSAIDs, likely due to structural and hydrophobicity differences influencing NSAID-membrane interactions and location within the membrane. NSAIDs primarily interacted with the lipid membrane through a combination of electrostatic interactions with the polar head group and hydrophobic interactions with the nonpolar alkyl tails. Aspirin and ibuprofen, with smaller molecular sizes and polar oxygen atoms, displayed stronger electrostatic interactions with the membrane's headgroup, while indomethacin, despite its charged form, was predominantly influenced by interactions with the hydrophobic tails owing to its larger nonpolar segment. Consequently, indomethacin encountered greater difficulty penetrating the membrane core compared to ibuprofen and aspirin, particularly in the ordered phase.

In the fluid phase, effects on membrane dynamics varied depending on the nature and concentration of the NSAIDs. Indomethacin suppressed membrane dynamics, whereas aspirin and ibuprofen slightly enhanced lateral diffusion at equivalent molar concentrations as shown in Fig. 10 (c). With increasing aspirin concentration, lateral diffusion coefficients saturated in the fluid phase, whereas higher concentrations of ibuprofen had the opposite effect, restricting lipid lateral diffusion[56, 97]. These observations underscore the importance of independently scrutinizing each NSAID when analyzing its mechanism of action. A deeper understanding of NSAID-membrane interactions holds promise for informing the development of more effective NSAIDs with reduced side effects.

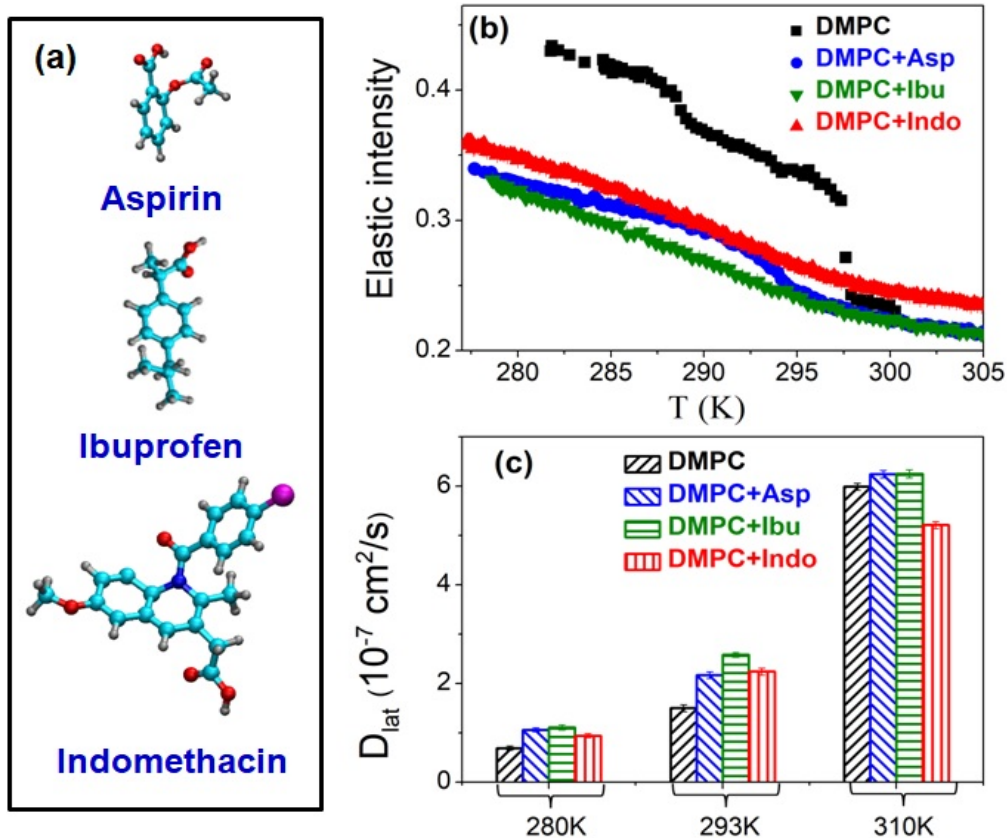

Fig. 10 (a) Chemical structures of aspirin, ibuprofen, and indomethacin. (b) $Q$ averaged elastic intensity for DMPC membrane, neat and with 25 mol% aspirin, ibuprofen or indomethacin in the heating cycle. (c) Lateral diffusion coefficient, $D_{lat}$ as obtained using QENS for the DMPC membrane, neat and with 25 mol% aspirin, ibuprofen or indomethacin at different temperatures. (adapted from refs.[56, 97])

To gain microscopic insights, we recently conducted MD simulations on DODAB lipid membranes both with and without aspirin, examining both ordered and fluid phases. Our simulations revealed that at 298 K and 310 K, the bilayers adopt interdigitated two-dimensional square phases, which become more rugged in the presence of aspirin. As temperature rises to 345 K, the bilayer transitions to a fluid state, with the disappearance of ripples. Aspirin tends to accumulate near oppositely charged headgroups, creating void spaces that increase interdigitation and order parameters. In the presence of aspirin, although the center of mass of lipids experiences structural arrest, they reach the diffusive regime faster, exhibiting higher lateral diffusion constants. These findings align with our QENS studies, indicating that aspirin functions as a plasticizer, enhancing lipid lateral diffusion in both ordered and fluid phases. Moreover, the addition of aspirin accelerates the relaxation time scales of bonds along the alkyl tails of DODAB due to increased lipid motions. Our results suggest that aspirin insertion is particularly favourable at physiological temperatures. Consequently, the DODAB bilayer, being ordered, stable, and exhibiting faster dynamics in the presence of aspirin, holds promise as a potential drug carrier for encapsulating aspirin, ensuring protected delivery, and facilitating targeted and controlled drug release, potentially with antibacterial activity in the future.

4. **Antioxidants & Stimulants**

Antioxidants encompass a range of compounds tasked with neutralizing free radicals and reactive oxygen species that inflict damage upon DNA, cell membranes, and various cellular components. They play pivotal roles in the development of numerous chronic diseases, including cardiovascular diseases, aging, cancer, and inflammation. α-tocopherol (αToc), the most biologically active form of vitamin E, serves as a crucial lipid-soluble antioxidant, intercepting free radicals diffusing into the membrane from the aqueous phase. Rather than being randomly distributed throughout the lipid membrane, αToc tends to aggregate in domains enriched with this compound. Numerous structural studies have been conducted to pinpoint the precise location of αToc in the lipid membrane. In most lipid environments, it is found near the lipid-water interface, with a perpendicular orientation placing the chromanol ring close to the polar lipid headgroup. However, an intriguing exception arises in the case of DMPC, where αToc is situated deep within the membrane, near the bilayer's center. Sharma et al.[42] have investigated effects of αToc on the dynamic and phase behavior of DMPC membranes. It was found that the inclusion of αToc had a profound impact on the phase behavior of the lipid membrane, notably suppressing the pre-transition and broadening the main phase transition. The increased broadening of the main phase transition suggests that the presence of αToc affects the cooperatively of the transition. Additionally, the main phase transition shifted towards lower temperatures, accompanied by a decrease in associated enthalpy, indicative of a smaller fraction of phospholipid molecules participating in the transition. QENS measurements revealed that αToc influences membrane dynamics. Both lateral and internal motions of lipids were observed to be modulated in the presence of αToc, suggesting a strong interaction between αToc and the lipid membrane. The alteration in membrane dynamics was found to be non-monotonous and dependent on the physical state of the membrane. Specifically, in the ordered phase, αToc acted as a plasticizer, enhancing both lateral and internal motions of the lipid membrane. However, in the fluid phase, it primarily restricted internal motion without significant effects on lateral motion.

Curcumin (diferuloylmethane), the primary compound in turmeric, has been investigated for its effects on membrane dynamics. Apart from its antioxidant properties, curcumin demonstrates a wide range of biological effects, including anti-inflammatory, anticancer, and wound-healing properties. Curcumin has been observed to influence the functionality and expression of a wide range of proteins, despite the absence of a definitive binding site for curcumin on any of these proteins. It is hypothesized that curcumin may modulate the functionality of membrane proteins indirectly by altering the physical properties of the host cell's membranes rather than through direct binding to proteins. This hypothesis has spurred investigations into the effects of curcumin on membrane dynamics. Sharma et al., have shown that curcumin significantly impacts the packing arrangement and conformations of DPPC lipids, resulting in enhanced membrane dynamics[78]. Specifically, a substantial acceleration of DPPC lateral motion in both the ordered and fluid phases in the presence of curcumin. The effects are particularly pronounced in the ordered phase, where the lateral diffusion coefficient increases by 23%, compared to a 9% increase in the fluid phase[78]. These findings provide crucial insights into the molecular mechanisms underlying the heightened lateral diffusion facilitated by curcumin. Interestingly, the localized internal motions of DPPC are minimally affected by curcumin, except for a slight enhancement observed in the ordered phase. When comparing the QENS results of different membrane-active compounds, Sharma et al., propose a hypothesis regarding their impact on lipid motion[78]. If a membrane-active compound adopts a surface-associated orientation, it is likely to influence the lateral motion of lipids. Conversely, if a membrane-active compound becomes deeply embedded within the membrane, it is expected to affect both the lateral and internal motions of the lipids. This distinction arises primarily due to the nature of lipid motion: lateral motion spans a long range within the membrane, whereas internal motion is confined to localized regions. This suggests that curcumin tends to localize preferentially at the membrane interface rather than adopting a transbilayer configuration. Furthermore, the unequivocal evidence demonstrating curcumin's modulation of membrane dynamics at a molecular level supports a potential action mechanism wherein curcumin acts as an allosteric regulator of membrane functionality. These insights contribute to our understanding of the mechanisms underlying curcumin's biological effects and its potential therapeutic applications.

Stimulants encompass a diverse range of substances known for boosting central nervous system activity. Widely utilized by a substantial portion of the population, these drugs serve various purposes, including enhancing performance, reaping medical benefits, and indulging in recreational activities. Prominent examples of stimulants comprise caffeine, nicotine, and cocaine. Among them, caffeine (1,3,7-trimethylxanthine) stands out as the most globally consumed legal psychoactive compound. Beyond its fatigue-fighting properties, caffeine has garnered attention for its potential antioxidative effects, potentially shielding against ailments like Alzheimer's disease. Various mechanisms have been proposed to elucidate caffeine's action, including the mobilization of intracellular calcium and the inhibition of adenosine and benzodiazepine ligand binding to neuronal membrane-bound receptors. Additionally, caffeine has been observed to interact with lipid membranes[113]. Sharma et al[113] explored the effects of caffeine on the microscopic dynamics of lipid membrane. It was shown that caffeine significantly alters the microscopic dynamics of the lipids within the system, with the observed effects contingent upon the lipid phase. Specifically, in the coagel phase, caffeine serves as a plasticizing agent, while in the fluid phase, it constrains both lateral and internal lipid motions. This study sheds light on how caffeine modulates membrane fluidity by influencing the dynamics of constituent lipids, a process intricately linked to the physical state of the bilayer.

5. **Depressants**

Depressants stand in contrast to stimulants as they work to decrease central nervous system activity, inducing relaxation, sedation, and diminished alertness. Common examples of depressants encompass alcohol, opioids, and similar substances. Rheinstädter and coworkers[114] have carried out a detailed investigation into the interaction between ethanol and DMPC membranes using neutron and X-ray scattering techniques. Their experiments involved DMPC hydrated powder with and without 5 wt% ethanol (equivalent to 2 mol%). It was found that ethanol molecules predominantly reside in the head group region of the bilayers, regardless of whether the membrane is in the gel or fluid phase. Furthermore, the presence of ethanol increases the permeability of the membrane in both phases. Neutron elastic intensity scans were conducted to elucidate the effects of ethanol on the phase behavior of the membrane. DMPC membranes undergo a pre-transition at 286 K and a main phase transition at 296.6 K. Ethanol was found to accentuate both transitions, making them more pronounced. Although the temperature of the main transition remained unchanged, the pre-transition was shifted by approximately 2.5 K towards higher temperatures in the presence of ethanol. Comparing the elastic intensity between pure DMPC and DMPC with ethanol, it was observed that ethanol did not alter the elastic intensity in the fluid phase.

However, in the gel phase, the presence of ethanol led to a drastic change in elastic intensity, indicating enhanced order in both the gel and ripple phases. The QENS experiments revealed that ethanol had a moderate effect on the lateral motion of lipids in the fluid phase, resulting in a slight decrease in lateral diffusion. In the gel phase, membranes exhibited a higher degree of order in the presence of ethanol, with lipid lateral diffusion slowing down by a factor of two. Overall, ethanol was found to induce a stiffer and more ordered structure in the ripple and gel phases of the membranes. Importantly, the dynamics of the membrane and hydration water were not significantly affected by the presence of ethanol in the physiologically relevant fluid phase at this alcohol concentration.

**6. Sterols**

Sterols constitute a vital lipid class pivotal in regulating numerous biological processes. Sterols play a critical role in facilitating the formation of liquid-ordered ($L_o$) membrane states, or lipid "rafts". These specialized membrane domains are believed to be essential for fundamental biological processes including signal transduction, cellular sorting, cytoskeleton reorganization, asymmetric growth, and the pathogenesis of infectious diseases. Sterols have been identified as key molecules responsible for maintaining membranes in a fluid state optimal for proper functionality. Notably, they play a significant role in regulating membrane thickness, facilitating the normal functioning of membrane proteins with diverse hydrophobic region lengths. Different kinds of sterols are found in nature such as cholesterol and ergosterol. Cholesterol predominates in vertebrates, whereas ergosterol is found in fungi and some protist. Chemical structure of ergosterol and cholesterol are shown in Fig. 11. Plants typically exhibit more intricate sterol compositions, featuring compounds like stigmasterol and sitosterol. These plant-specific sterols play pivotal roles in embryonic plant growth.

**6.1 Cholesterol**

Cholesterol comprises a small polar hydrophilic hydroxyl group and a relatively larger hydrophobic steroid ring, lacking inherent ability to self-assemble into distinct structures. Its multifunctional role within cells encompasses modulating various physical properties of the plasma membrane, including mechanical strength (bending and compressibility modules) and fluidity. Extensive investigations into the interaction between cholesterol and lipid membranes have yielded a plethora of intriguing findings. These include condensing effects, reduction in membrane permeability, suppression of the main phase transition of lipids at sufficiently high concentrations, augmentation of acyl chain ordering in the fluid phase while diminishing ordering in the gel phase, among others[115]. Busch et al.[116] conducted QENS experiments on DMPC membranes with varying concentrations of cholesterol, employing two different energy resolutions: 60 µeV and 4 µeV, corresponding to observation time scales of 60 ps and 900 ps, respectively. Their findings indicated that above the main phase transition, the inclusion of cholesterol resulted in a reduction in quasielastic broadening, indicative of constrained membrane dynamics. This effect was particularly pronounced at higher energy resolutions (4 µeV) or longer time scales (900 ps), where membrane dynamics exhibited a monotonous decrease with increasing cholesterol concentration. Below the main phase transition, however, the addition of cholesterol did not alter membrane dynamics. In contrast, Sharma et al.[48] conducted a comprehensive investigation into the impact of cholesterol on the lateral motion of the membrane. They found that cholesterol significantly restricted lateral motion, particularly above the main phase transition temperature. At 310 K, the addition of 20 mol% cholesterol reduced the lateral diffusion coefficient from $7.7 \times 10^{-7}$ cm$^2$/s to $2.8 \times 10^{-7}$ cm$^2$/s. The main effect of cholesterol is to decrease the available free area and observed results can be explained based on Eq. (22), free area theory. This restriction can be attributed to cholesterol's ability to promote tight packing of lipid molecules resulting in reduced average area per phospholipid, $a(T)$. Moreover, one also needs to subtract area occupied by cholesterol from $a(T)$ to obtain available free area. [117]. At 280 K, no significant effect of cholesterol on the lateral motion of DMPC was observed, with a lateral diffusion coefficient of $0.7 \times 10^{-7}$ cm$^2$/s.

**6.2 Ergosterol**

Fungal infections pose a growing threat to global health, fueled by factors such as climate change and immunocompromised populations. Ergosterol plays a pivotal role in the structure and function of fungal cell membranes, serving as a crucial component that influences membrane fluidity, integrity, and permeability. Moreover, the distinct interactions of ergosterol with therapeutic agents underscore its potential as a target for developing novel antifungal strategies. For example, ergosterol uniquely interacts with amphotericin B, a potent antifungal drug, distinguishing it from cholesterol and suggesting its potential as a specific therapeutic target. Due to similar chemical structure, ergosterol is thought to modulate lipid membrane in similar manner as cholesterol. Recent study has unveiled fundamental differences in the effects of ergosterol and cholesterol on the structure and dynamics of lipid membranes[118]. Unlike cholesterol, ergosterol is embedded more shallowly in the lipid bilayer and exhibits a smaller condensation effect, resulting in less significant changes to membrane thickness. Moreover, ergosterol can both rigidify and soften membranes, depending on its concentration, contrasting with cholesterol's consistent role in increasing membrane rigidity. QENS experiments on 1-palmitoyl-2-oleoyl-sn-glycero-3-phosphocholine (POPC) lipid membranes in the absence and presence of ergosterol and cholesterol have revealed distinct diffusion behaviors. Variations of HWHMs of Lorentzian function associated with lateral diffusion with $Q^2$ for POPC membrane, neat and in presence of ergosterol and cholesterol are shown in Fig. 11. While lateral diffusion in POPC and POPC with cholesterol follows continuous diffusion, POPC with ergosterol requires jump diffusion to describe the observed data. The lateral diffusion coefficients were measured as $5.3 \times 10^{-7}$ cm$^2$/s for POPC, $2.9 \times 10^{-7}$ cm$^2$/s for POPC with cholesterol, and $3.8 \times 10^{-7}$ cm$^2$/s for POPC with ergosterol. Interestingly, in the fluid phase, cholesterol does not alter the continuous Fickian diffusion seen in POPC but significantly reduces its diffusion rate.

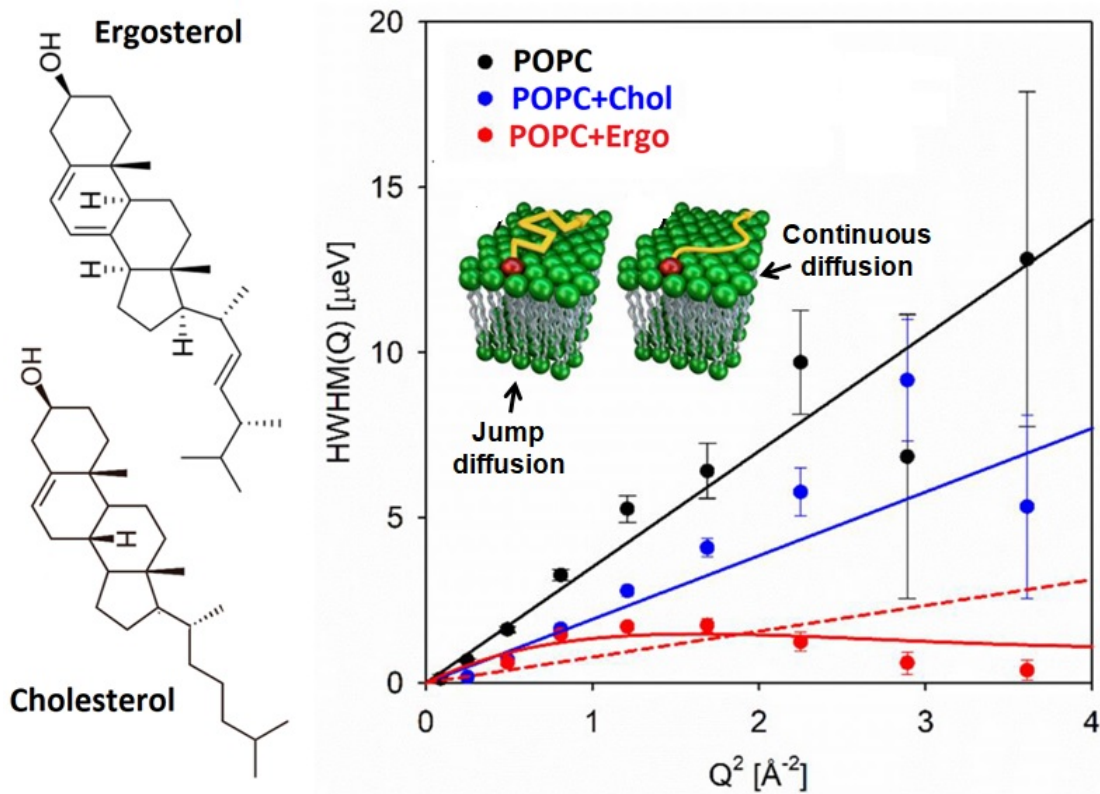

**Fig. 11** Chemical structure of ergosterol and cholesterol. Variation of half width at half maxima (HWHM) of Lorentzian function associated with lateral diffusion with $Q^2$ for POPC membrane, neat and in presence of ergosterol and cholesterol. The best fits (solid lines) are given by the continuous Fickian diffusion model for POPC and POPC+cholesterol and jump diffusion model for POPC+ergosterol. For direct comparison, fits using continuous Fickian diffusion for POPC+ergosterol is shown by a dashed line (adapted from[69]).

Conversely, ergosterol's presence results in a moderate reduction in diffusion rate and transform the mechanism of POPC dynamics to a distinct jump diffusion process. These findings suggest distinct interactions of ergosterol with membranes compared to cholesterol, highlighting the unique role of ergosterol in membrane dynamics.

## 7. Ionic Liquids

Ionic liquids (ILs) represent a distinctive class of organic salts characterized by their low melting point below 100 °C and unique physicochemical properties. ILs are widely applied across diverse fields, including bio-nanotechnology, where they play pivotal roles in solubilizing drugs, stabilizing biomolecules, and exhibiting antibacterial properties. Despite their broad utility, the precise mechanism underlying their antimicrobial activity remains elusive, impeding their optimized application in bio-nanomedicine. Bakshi et al., have unveiled that ILs induce necrosis-dependent cell death through membrane damage, prompting an exploration into their interaction with model cell membranes[119]. Our findings elucidate how IL incorporation modulates the structure, dynamics, and phase behavior of lipid membranes[38,50,99,100,119-124]. Our investigation delved into the effects of different ILs, varying in alkyl chain length and anions, on various lipid membranes, including zwitterionic DMPC, DPPC, sphingomyelin, cationic DHDAB, and liver lipid extract[38,50,99,100,119-124]. Utilizing X-ray reflectivity, we reveal the formation of interdigitated domains upon IL addition[122]. IL also exerts a notable influence on the phase behavior of lipid membranes. They suppress the pre-transition and shift the main transition temperature downward. Additionally, the main phase transition becomes broader, indicating a decrease in the cooperativity of the phase transition. Despite the observed shift in $T_m$ towards lower temperatures and the broadening of the peak, the addition of ILs does not significantly affect the enthalpy (ΔH) associated with the main transition, suggesting no change in the number of lipids participating in the main phase transition. Furthermore, we observe that the interaction between ILs and the membrane depends on the alkyl chain length of IL[50,99,121]. As the alkyl chain length of IL increases, the binding affinity of IL with the lipid membrane also increases, indicating that hydrophobic interactions between IL and the membrane play a crucial role. These results align with dye leakage assay measurements, which suggest that the stronger binding of DMIM[Br] IL correlates with higher membrane permeability[50]. Across all cases, ILs functioned as plasticizers, augmenting membrane dynamics, with their effects contingent on alkyl chain length, which intensified with longer chains. ILs accelerated lateral diffusion in both ordered and fluid phases, with more pronounced effects observed in the ordered phase[50,99]. For instance, the inclusion of BMIM[BF4] increased lateral diffusion of DMPC by 22% in the ordered phase, compared to just 2% in the fluid phase[99]. Longer chain ILs displayed enhanced plasticizing effects, highlighting the importance of hydrophobic interactions in their mechanisms of action. For example, in the fluid phase of DMPC, C10MIM[BF4] boosted lateral diffusion by 38%, while C4MIM[BF4] exhibited a modest 2% increase[99]. This underscores the significant role of hydrophobic interactions between ILs and lipid membranes. The stronger impact of longer chain length ILs on membrane dynamics correlates with heightened antimicrobial efficacy, as demonstrated by a decrease in MIC, emphasizing the pivotal role of membrane dynamics in influencing membrane fluidity and permeability, thereby impacting cell stability[50]. This augmentation in membrane dynamics led to enhanced membrane fluidity and permeability, confirmed through dye leakage and flow cytometry assays, establishing a direct relationship between IL-induced membrane dynamics and antimicrobial activity. These findings underscore the tunability of IL antibacterial properties through the manipulation of alkyl chain length, highlighting the significance of comprehending their effects on membrane dynamics for tailored bio-nanomedicine applications. Interestingly, ILs not only accelerated lateral diffusion in single lipid systems but also in complex liver lipid extracts[100], mimicking cell membranes more closely. This universality of the plasticizing effect suggests that it is independent of membrane composition. QENS data provided a quantitative framework for elucidating the effects of ILs on the dynamical and phase behavior of model cell membranes, essential for understanding their action mechanisms comprehensively.

This study also holds broader implications for the development of efficient drug delivery systems. Liposomes have emerged as highly promising carriers for drug delivery, with the stability and release kinetics of liposomes crucially dependent on the phase behavior and membrane fluidity. Our findings highlight the pivotal role of IL incorporation in modulating both the phase behavior and fluidity of the membrane, which in turn influences the balance of electrostatic and hydrophobic interactions and enhances the efficiency of cargo transport. This insight underscores the potential of IL-modified liposomes as effective platforms for advanced drug delivery systems, offering opportunities for improved therapeutic outcomes and targeted delivery strategies.

The impact of DMIM[Br] on the structural, dynamical, and phase behavior of cationic DHDAB vesicles has been thoroughly investigated[120]. In the heating cycle, pure DHDAB membranes exhibit two distinct endothermic transitions: the pre-transition occurring at 303 K and the main transition at 318 K[125]. The pre-transition marks a solid-to-solid polymorphic shift from the coagel phase to the intermediate crystalline (IC) phase, accompanied by mild disorder in alkyl chains. Conversely, the main transition signifies the transformation from the IC phase to the fluid phase, characterized by decreased packing density, increased head group hydration, and heightened alkyl chain disorder. Notably, a significant hysteresis is observed in the phase behavior during the cooling cycle, with direct transformation from the fluid to coagel phase. Incorporation of DMIM[Br] exerts concentration-dependent effects on DHDAB bilayer phase behavior[120]. At 10 and 25 mol %, DMIM[Br] eliminates the pre-transition and lowers the onset temperature of the main transition, while inducing an intermediate gel phase formation during the cooling cycle, akin to observations in DODAB membranes. This phenomenon is attributed to strengthened hydrophobic attraction between tails and enhanced electrostatic repulsion upon DMIM[Br] insertion. However, at 40 mol %, the formation of the intermediate gel phase is suppressed, likely due to self-aggregation of DMIM[Br] into micellar forms, altering the interaction landscape. QENS data reveal enhanced lipid mobility in coagel and fluid phases upon DMIM[Br] addition, suggesting its role as a plasticizer, thereby enhancing membrane fluidity across all phases. These findings underscore the modulation of membrane phase behavior and fluidity by DMIM[Br], pivotal for efficient cargo transport, through control of electrostatic and hydrophobic interactions.

## 8. Unsaturated lipids (e.g. monoolein)

Ensuring the flexibility of cell membranes is crucial for numerous cellular processes. In organisms like plants and cyanobacteria, maintaining membrane fluidity becomes particularly vital in colder environments, where fluidity naturally decreases. These organisms counter this by adjusting the composition of their membrane lipids, increasing the number of double bonds in fatty acids and thereby the number of unsaturated lipids in the membrane. Investigating how unsaturated lipids contribute to this maintenance is therefore valuable. Singh et al., explored the impact of monoolein (MO), a significant unsaturated lipid, on the behavior and dynamics of cationic DODAB membranes[53]. MO stands out for its non-toxic, biodegradable, and biocompatible properties, making it widely applicable in drug delivery, emulsion stabilization, and protein crystallization. Notably, liposomes formed from DODAB:MO combinations exhibit promise as carriers for gene therapy. These formulations demonstrate reduced toxicity and effectively facilitate cell transfection in in-vitro studies. Our neutron elastic intensity scan revealed significant alterations in the phase behavior of DODAB membranes upon incorporation of MO, evident in both heating and cooling cycles. During heating, the transition from the coagel to fluid phase becomes less abrupt and more diffuse, with the phase transition temperature shifting towards lower values. Conversely, in the cooling cycle, the formation of an intermediate gel phase is suppressed, leading DODAB to transition directly from the fluid phase to the coagel phase. This suggests a synchronized ordering of lipid headgroups and tails, contrasting with the nonsynchronous changes observed in pure DODAB membranes. Furthermore, our investigation into membrane dynamics unveiled notable effects of MO. In the coagel phase, MO serves as a plasticizer, enhancing quasielastic broadening and thus increasing membrane flexibility. This disrupts the tightly packed lipid structure characteristic of the coagel phase. Conversely, in the fluid phase, MO acts as a stiffening agent, restricting lateral diffusion of lipid molecules. The lateral diffusion coefficient decreases from $22 \times 10^{-6}$ cm$^2$/s to $17 \times 10^{-7}$ cm$^2$/s in DODAB membranes with MO, attributed to the reduction in available free area per lipid molecule[53]. Although the nature of internal motions within the fluid phase remains largely unchanged with MO addition, there is a reduction in the HWHM corresponding to segmental motion, indicating a deceleration of segmental motions.

Our research also highlights the pivotal role of the location of guest molecules within the lipid membrane in dictating their effects on membrane dynamics. Alongside factors such as size and the interaction with the membrane, the location of these molecules significantly influences their impact on membrane dynamical behavior. When situated at the interface between the lipid and aqueous phase, guest molecules predominantly influence lateral motion, leaving internal motion largely unaffected. Conversely, molecules residing deep within the hydrophobic core of the membrane have the potential to influence both lateral and internal motions. AMP provides a pertinent example: at low concentrations, they bind to the surface of the lipid membrane, primarily affecting lateral motion due to the diffusion of lipids within the leaflet. In contrast, compounds like ILs, NSAIDs penetrate deep into the membrane, influencing both lateral and internal motions. Our investigations have revealed that the incorporation of ILs induces disorder within the membrane, thereby enhancing both lateral and internal motion. Furthermore, the behavior of cholesterol within the membrane contrasts with that of ILs: while cholesterol penetrates into the membrane core, it restricts both lateral and internal motions of lipids in the fluidic phase of the membrane. This discrepancy can be attributed to the differing effects of cholesterol and ILs on membrane orderliness: cholesterol enhances membrane order, whereas ILs induce disorder. These findings underscore the sensitivity of membrane dynamics to the presence of membrane-active molecules. Understanding these interactions not only sheds light on fundamental aspects of membrane biology but also holds potential for informing the design of therapeutic agents and elucidating their effects on cellular membranes. Results on effects of these membrane active compounds on the membrane dynamics is summarised in Table-2.

**Table-2** Effects of Membrane-Active Compounds on Lipid Lateral Diffusion Coefficients in Model Membrane Systems: Experimental Conditions and Key Findings.

| S. No. | Compound and concentration | Membrane system | Spectrometer (Resolution) | T (Phase) | Con. (mol %) | $D_{lat}$ ($10^{-7}$ cm$^2$/s) | Main findings | Refs. |
|---|---|---|---|---|---|---|---|---|
| 1. | **Membrane Active Peptides** | | | | | | | |
| 1.1 | **Amyloid peptides: Amyloid Beta (Aβ)** | | | | | | | |
| 1.1.1 | Aβ (25-35) 3 mol % | DMPC/DMPS (92:8) Supported lipid multilayar | NEAT (93 μeV) | 290 K (Ordered) | 0 | 6 | (i) Aβ (25-35) accelerates lateral diffusion in both ordered & fluid phases (ii) Aβ (25-35) also enhances out of plane motion of lipid. | 105 |
|  |  |  |  |  | 3 | 8 |  |  |
|  |  |  |  | 320 K (Fluid) | 0 | 18 |  |  |
|  |  |  |  |  | 3 | 24 |  |  |
| 1.1.2 | Aβ (22-40) 3 mol % | DMPC/DMPS (92:8) Supported lipid multilayar | NEAT (93 μeV) | 320 K (Fluid) | | - | Aβ (22-40) accelerates lateral diffusion of lipid in fluid phase | 106 |
|  | Aβ (22-40) 1.5 mol % | DMPC/DMPS (92:8) Supported lipid bilayer | IN16 (1 μeV); IN5 (15 μeV); TOFTOF (100 μeV, 25 μeV) | 288 K (Ordered) | 0 | 1.02 | In the ordered phase, Aβ (22-40) enhances lateral diffusion. Near the main phase transition temperature (303 K), the trend is reversed, Aβ (22-40) causes decrease in the lateral diffusion coefficient ($D_{lat}$). | 88 |
|  |  |  |  |  | 1.5 | 1.17 |  |  |
|  |  |  |  | 303 K (near the phase transition) | 0 | 2.08 |  |  |
|  |  |  |  |  | 1.5 | 1.58 |  |  |
| 1.1.3 | Aβ (1-40) 3.3 mol % | DMPG 5 wt % unilamellar vesicles (ULVs) | BASIS (3.4 μeV) | 280 K (Ordered) | 0 | 1.1 | Aβ (1-40) mainly affects the long range lateral motion of lipid molecules, especially in the fluid phase, but not the localised internal motion. | 43 |
|  |  |  |  |  | 3.3 | 0.9 |  |  |
|  |  |  |  | 310 K (Fluid) | 0 | 5.5 |  |  |
|  |  |  |  |  | 3.3 | 6.9 |  |  |

| | | | | | | | |
|---|---|---|---|---|---|---|---|
| 1.2 | **Antimicrobial peptides** | | | | | | |
| 1.2.1 | Melittin 0.2 mol % | DMPC 5 wt % ULVs | BASIS (3.4 µeV) | 280 K (Ordered) | 0 | 0.7 | (i) In the ordered phase, melittin acts as a plasticizer, enhancing the lateral motion of lipid. However, in the fluid phase it acts as a stiffening agent, restricting the lateral motion of the lipid.<br><br>(ii) No significant effect of melittin on $D_{lat}$ in the presence of cholesterol<br><br>(iii) Destabilizing effect of melittin on membranes can be mitigated by the presence of cholesterol | 48 |
| | | | | | 0.2 | 1.1 | | |
| | | | | 310 K (Fluid) | 0 | 7.7 | | |
| | | | | | 0.2 | 2.4 | | |
| | | DMPC with 20 mol % cholesterol 5 wt % ULVs | BASIS (3.4 µeV) | 280 K | 0 | 0.7 | | |
| | | | | | 0.2 | 0.7 | | |
| | | | | 310 K | 0 | 2.8 | | |
| | | | | | 0.2 | 2.8 | | |
| 1.2.2 | Alamethicin 0.5 mol % | DMPC 7 wt % ULVs | BASIS (3.4 µeV) | 280 K (Ordered) | 0 | 0.7 | (i) In the ordered phase, no significant effect of alamethicin on $D_{lat}$<br><br>(ii) In the fluid phase, alamethicin restricts lateral diffusion of lipid.<br><br>(iii) Effect of cationic (+5 e) melittin on lateral diffusion and phase behavior of membrane are stronger than anionic (-1e) alamethicin.<br><br>(iv) Proposed a new action mechanism of AMPs in which by modulating lateral motion at very low concentration, AMPs kills the bacterial cells | 98 |
| | | | | | 0.5 | 0.7 | | |
| | | | | 310 K (Fluid) | 0 | 5.0 | | |
| | | | | | 0.5 | 3.9 | | |
| 1.2.3 | Aurein Different molar concentrations (0, 1, 1.7 & 2.5 mol %) | DMPC 7 wt % ULVs | BASIS (3.4 µeV) | 280 K (ordered) | 0 | 0.7 | (i) In the ordered phase, no significant effect on $D_{lat}$.<br><br>(ii) In the fluid phase, $D_{lat}$ decreases monotonously with increase in concentration of aurein.<br><br>(iii) This decrease in lateral diffusion leads to lateral segregation of lipids.<br><br>(iv) Melittin remains as an exception which acts plasticizer in the ordered phase and acts oppositely in the fluid phase. Other AMPs (alamethicin, aurein) though restricts lateral motion in the | 91 |
| | | | | | 1.7 | 0.8 | | |
| | | | | 293 K (ordered) | 0 | 1.6 | | |
| | | | | | 1.7 | 1.5 | | |
| | | | | 310 K (fluid) | 0 | 6.0 | | |
| | | | | | 1.7 | 4.7 | | |
| | | | | 320 K (Fluid) | 0 | 8.7 | | |
| | | | | | 1 | 7.5 | | |
| | | | | | 1.7 | 6.1 | | |
| | | | | | 2. | 5.4 | | |

| | | | | | | | |
|---|---|---|---|---|---|---|---|
| | | | | | | fluid phase but does not affect dynamics much in the ordered phase. | |
| 1.2.4 | **Ubiquicidin (UBI)** | | | | | | |
| 1.2.4.1 | UBI (31-38) 4 mol % | DPPG 5 wt % ULVs | IRIS (17µeV) | 330 K (Fluid) | 0 | 17.4 | UBI (31-38) restricts lateral diffusion of DPPG. | 27 |
| | | | | | 4 | 12.5 | | |
| 1.2.4.2 | UBI (29-41) 0, 2 & 4 mol % | DPPG 5 wt % ULVs | IRIS (17 µeV) | 330 K (Fluid) | 0 | 17.4 | (i) UBI (29-41) restricts lateral diffusion of DPPG. (ii) $D_{lat}$ decreases monotonously with increase in concentration of UBI (29-41). (iii) UBI (29-41) acts stronger than UBI (31-38) | |
| | | | | | 2 | 13.8 | | |
| | | | | | 4 | 10.5 | | |
| 2. | **Transmembrane Peptide** | | | | | | | |
| 2.1 | Transferrin receptor (TFRC) protein | DMPC 10 wt % ULVs | IN16 B (0.75 ueV) | 310 K (Fluid) | 0 | 2.1 | TFRC protein restricts lateral diffusion of the lipid. | 44 |
| | | | | | 6 | 0.8 | | |
| 3, | **NSAIDs** | | | | | | | |
| 3.1 | Aspirin  0, 25 and 50 mol % | DMPC 7 wt % ULVs | BASIS (3.4 µeV) | 280 K (Ordered) | 0 | 0.7 | (i) In the ordered phase, $D_{lat}$ increases monotonously with increase in concentration of aspirin. (ii) Effects of aspirin on membrane dynamics are more prominent near the main phase transition temperature, as main phase transition gets broaden and shifts towards the lower temperature. (iii) In the fluid phase, at lower concentration, addition of aspirin enhances $D_{lat}$ then it saturates | 56 |
| | | | | | 25 | 1.1 | | |
| | | | | | 50 | 1.2 | | |
| | | | | 293 K (Ordered) | 0 | 1.5 | | |
| | | | | | 25 | 2.2 | | |
| | | | | | 50 | 2.5 | | |
| | | | | 310 K (Fluid) | 0 | 6.0 | | |
| | | | | | 25 | 6.3 | | |
| | | | | | 50 | 6.3 | | |
| | | DODAB 70 mM vesicles | IRIS (17µeV) | 330 K (Fluid) | 0 | 22 | In the fluid phase, $D_{lat}$ of DODAB increases slightly with 25 mol % aspirin | 110 |
| | | | | | 25 | 24 | | |
| 3.2 | Ibuprofen | DMPC | BASIS (3.4 µeV) | 280 K (Ordered) | 0 | 0.7 | (i) In the ordered phase, $D_{lat}$ increases monotonously with | 97 |
| | | | | | 25 | 1.1 | | |

| | | | | | | | |
|---|---|---|---|---|---|---|---|
| | 0, 25 and 50 mol % | 7 wt % ULVs | | | 50 | 1.6 | increase in concentration of ibuprofen.<br><br>(ii) In fluid phase, $D_{lat}$ has non-monotonous dependence on concentration: first increases at 25 mol % then it decreases to 50 mol % ibuprofen |
| | | | | 293 K (Ordered) | 0 | 1.5 | |
| | | | | | 25 | 2.6 | |
| | | | | | 50 | 3.3 | |
| | | | | 310 K (Fluid) | 0 | 6.0 | |
| | | | | | 25 | 6.3 | |
| | | | | | 50 | 5.4 | |
| 3.3 | Indomethacin<br><br>25 mol % | DMPC<br><br>7 wt % ULVs | BASIS (3.4µeV) | 280K (Ordered) | 0 | 0.7 | (i) In the ordered phase, all NSAIDs generally enhances the membrane fluidity. Among these ibuprofen affects dynamics most compared to other suggesting stronger interaction with lipid.<br><br>(ii) In the fluid phase, 25 mol % indomethacin restricts lateral diffusion which is in contrast with the action of aspirin & ibuprofen. This suggests that effects of NSAIDs on $D_{lat}$ depends on the concentration and type of NSAIDs. For ex. ibuprofen at 25 mol % enhances but at 50 mol % restricts lateral diffusion of lipids. | 97 |
| | | | | | 25 | 0.9 | |
| | | | | 293K (Ordered) | 0 | 1.5 | |
| | | | | | 25 | 2.2 | |
| | | | | 310 K (Fluid) | 0 | 6.0 | |
| | | | | | 25 | 5.2 | |
| | | DODAB<br><br>70 mM vesicles | IRIS (17.5µeV) | 330K (Fluid) | 0 | 22 | In the fluid phase, incorporation of indomethacin restricts lateral diffusion of DODAB. A very similar action was observed for zwitterionic DMPC membrane. | 110 |
| | | | | | 25 | 15 | |
| **4.** | **Antioxidants & Stimulants** | | | | | | |
| 4.1 | Vitamin E<br><br>α-tocopherol (aToc) | DMPC<br><br>7 wt % ULVs | BASIS (3.4 µeV) | 280 K (Ordered) | 0 | 0.7 | In the ordered phase, aToc enhances both lateral & internal motion. However, in the fluid phase, aToc restricts only internal motion, no significant effect on lateral motion was observed | 42 |
| | | | | | 10 | 0.8 | |
| | | | | | 20 | 0.8 | |
| | | | | 293 K (Ordered) | 0 | 1.4 | |
| | | | | | 10 | 1.6 | |
| | | | | | 20 | 1.9 | |
| | | | | 310 K (Fluid) | 0 | 5 | |
| | | | | | 10 | 4.9 | |
| | | | | | 20 | 5 | |
| 4.2 | Curcumin | DPPC<br><br>5 wt % ULVs | IRIS (17 µeV) | 310 K (Ordered) | 0 | 5.0 | (i) In the ordered and fluid phase, curcumin enhances lateral diffusion about 23 % | 78 |
| | | | | | 2 | 6.0 | |
| | | | | 330 K | 0 | 17.7 | |
| | 2 mol % | | | (Fluid) | 2 | 20.6 | and 9 % respectively.<br><br>(ii) Effects are more prominent in the ordered phase than in the fluid phase<br><br>(iii) Curcumin act as an allosteric regulator . | |
| 4.3 | Caffeine<br><br>25 mol % | DODAB<br><br>70 mM vesicles | IRIS<br>(17.5 µeV) | 330 K<br>(Fluid) | 0<br>25 | 22<br>12 | (i) In the coagel phase, caffeine acts as a plasticizing agent, whereas in the fluid phase, it restricts the lateral and internal motions of the lipids. | 113 |
| 5. | **Depressants** | | | | | | | |
| 5.1 | Ethanol | DMPC<br><br>Hydrated powder | HFBS<br>(0.8 µeV) | 286 K (ordered) | 0<br>2 | 0.9<br>0.5 | (i) In the ordered phase, lateral diffusion in the presence of ethanol is slowed down by a factor of 2 as compared to pure DMPC.<br><br>(ii) Ethanol molecules reside in the head group region of the bilayers and enhances permeability of membrane | 114 |
| | | | | 308 K (Fluid) | 0<br>2 | 5.2<br>4.8 | | |
| 6. | **Sterols** | | | | | | | |
| 6.1 | Cholesterol | DMPC<br><br>5 wt % ULVs | BASIS<br>(3.4 µeV) | 280 K (ordered) | 0<br>20 | 0.7<br>0.7 | Cholesterol restricts lateral diffusion of lipid in the fluid phase. | 48 |
| | | | | 310 K (Fluid) | 0<br>20 | 7.7<br>2.8 | | |
| | | DMPC<br><br>50 wt % Hydrated lipid powder | TOFTOF (60 µeV)<br><br>TOFTOF (4 µeV) | 293 K (ordered)<br>303 K<br>313 K (fluid) | 0 to 40 | -<br>-<br>- | In the fluid phase, incorporation of cholesterol leads to decrease the mobility of lipid with increasing the cholesterol content. Effects are more prominent at high energy resolution (4 µeV) or longer observational time. | 116 |
| | | POPC<br><br>ULVs | BASIS<br>(3.7 µeV) | 303 K (fluid) | 0<br>30 | 5.3<br>2.9 | In the fluid phase, the addition of cholesterol significantly reduces the lateral diffusion coefficient but it does not change the nature of diffusion mechanism which is continuous Fickian diffusion | 118 |
| 6.2 | Ergosterol | POPC<br><br>ULVs | BASIS<br>(3.7 µeV) | 303 K (fluid) | 0<br>30 | 5.3<br>3.8[a] | [a] *jump diffusion*<br><br>Ergosterol restricts the lateral diffusion moderately but | |
| | | | | | | | change the nature of the diffusion process. A jump diffusion is observed in the presence of Ergosterol | |
| 7. | **Ionic Liquids (ILs)** | | | | | | | |
| 7.1 | BMIM[BF4] 100 mM | DMPC ULVs | DNA (3.6 µeV) | 283 K (Ordered) | 0 | 0.9 | (i) In both ordered and fluid phases, IL accelerates the lateral diffusion. (ii) Effects are stronger in the ordered phase. (iii) Longer chain IL is found to be stronger plasticizer suggesting hydrophobic interaction between IL-lipid membrane play an important role in their action mechanism. | 99 |
| | | | | | 20 | 1.1 | | |
| | | | | 293 K (Ordered) | 0 | 1.6 | | |
| | | | | | 20 | 1.7 | | |
| | | | | 303 K (Fluid) | 0 | 4.8 | | |
| | | | | | 20 | 4.9 | | |
| 7.2 | DMIM[BF4] 100 mM | DMPC ULVs | DNA (3.6 µeV) | 283 K (Ordered) | 0 | 0.9 | | |
| | | | | | 20 | 1.3 | | |
| | | | | 293 K (Ordered) | 0 | 1.6 | | |
| | | | | | 20 | 3.6 | | |
| | | | | 303 K (Fluid) | 0 | 4.8 | | |
| | | | | | 20 | 6.6 | | |
| | | Liver lipid extract 5 wt % ULVs | IRIS (17.5 µeV) | 310 K | 0 | 11.6 | [b]: wt percentage ILs accelerate lateral diffusion not only for single lipid system but also for a complex liver lipid extract which is a mixture of various lipid and closer mimic to cell membrane | 100 |
| | | | | | 10[b] | 13.3 | | |
| | | | | 330 K | 0 | 18.2 | | |
| | | | | | 10[b] | 21.5 | | |
| **7.3** | DMIM[Br] | DHDAB 70 mM Vesicles | IRIS (17.5 µeV) | 330 K (Fluid) | 0 | 27 | Incorporation of IL enhances membrane dynamics in both ordered & fluid phases, suggesting that DMIM[Br] acts as a plasticizer. | 119 |
| | | | | | 25 | 30 | | |
| | | DPPC 5 wt % ULVs | IRIS (17.5 µeV) | 330 K (fluid) | 0 | 16.8 | With increase in concentration of IL, $D_{lat}$ increases Increase in lateral diffusion ultimately leads to increased permeability and subsequent bacterial death. | 50 |
| | | | | | 25 | 21.3 | | |
| | | | | | 50 | 24.4 | | |
| **7.4** | **HMIM[Br]** | DPPC 5 wt % ULVs | IRIS (17.5 µeV) | 330 K (fluid) | 0 | 16.8 | | 50 |
| | | | | | 25 | 19.3 | | |
| **8.** | **Unsaturated lipids** | | | | | | | |
| 8.1 | Monoolein (MO) | DODAB 70 mM vesicles | IRIS (17.5 µeV) | 330 K (fluid) | 0 | 22 | In the fluid phase, MO restricts both lateral and internal motions of the lipid | 53 |
| | | | | | 33 | 17 | | |

DMPC: Zwitterionic, $T_m$=297 K
DPPC: Zwitterionic, $T_m$ =314 K
DMPS: Anionic, $T_m$ =309 K
DMPG: Anionic, $T_m$: 297 K
DPPG: Anionic, $T_m$ =314 K
DHDAB: Cationic $T_m$ = 318 K
DODAB: Cationic, $T_m$ = 327 K

**CONCLUDING REMARKS AND FUTURE DIRECTIONS**

This article provides an in-depth exploration of the lateral diffusion of lipids within homogeneous membrane systems, highlighting the complexities and sources of heterogeneity that influence this process. Strong emphasis is given to the discussion on significant variability in the lateral diffusion coefficient ($D_{lat}$) across different spatial and temporal scales. In particular, the measurement of $D_{lat}$ at the mesoscopic length scale are essential to reconcile the discrepancies in diffusion coefficients obtained from neutron scattering (microscopic) and macroscopic experimental techniques (fluorescence, PFG-NMR, etc.). Incoherent neutron spin echo (NSE) shows promise in bridging this gap, with early experiments indicating its potential to deepen our understanding of lipid diffusion at mesoscopic length scales. Moreover, molecular dynamics (MD) simulations and preliminary quasielastic neutron scattering (QENS) experiments suggest the presence of sub-diffusive motion in lipid lateral, indicating a need for a comprehensive diffusion model which across different length scales. This model should address the transitions from continuous to flow-like motion at higher $Q$-values and incorporate the sub-diffusive behavior noted in MD simulations. Integrating the results from NSE and QENS with macroscopic experiments can potentially provide sufficient data for building such a cohesive model of lipid lateral diffusion. Employing these models can then provide consistent and reliable lateral diffusion coefficients.

Several factors influencing lateral diffusion are discussed in detail, including area per lipid, hydration, membrane curvature, temperature, and pressure. We have also explored the impact of membrane-active compounds on $D_{lat}$ and the various factors influencing the interaction between these compounds and lipid membranes. Key considerations include the size, charge, and location of these compounds, as well as the polarity and composition of the membrane. Understanding these elements is critical for elucidating the mechanisms of action of these compounds in biological systems.

The majority of studies on lipid diffusion have focused on simplified systems, often examining single-component lipid membranes to uncover fundamental mechanisms. However, cell membranes are far more complex, comprising intricate mixtures of lipids, proteins, and various small molecules. This heterogeneity within cell membranes varies across organelles, cell types, organisms, and even in different physiological or pathological states. Various factors such as physical obstructions, binding interactions, the influence of the membrane skeleton, and biological regulation contribute to the complex patterns of diffusion in the plasma membrane. To fully understand these patterns, a comprehensive approach that considers the impact of these factors across different spatial and temporal scales is required.

Moreover, it's essential to analyze obstacles that impede diffusion within the membrane, whether they arise from proteins or other structural components, resulting in heterogeneous environments where movement is constrained. Binding interactions between membrane components, such as lipids and proteins, also play a significant role, potentially forming microdomains or clusters that affect the overall diffusion landscape. Additionally, the membrane skeleton, a network of cytoskeletal elements beneath the membrane, can create barriers or scaffolds that alter diffusion pathways. Biological regulation, which includes cellular signaling and response mechanisms, can further influence the mobility and organization of membrane components.

To gain a more comprehensive understanding of lipid diffusion, it's critical to shift the research focus toward more complex systems that reflect this inherent heterogeneity. This includes exploring membranes formed from natural lipid extracts and studying in vivo membrane dynamics. By doing so, one can better understand how this complexity impacts diffusion processes. Advances in neutron spectrometry, selective deuteration techniques, and sophisticated measurement and analysis methods are making these studies increasingly viable. For example, recent research efforts : Sharma et al.[100] explored the effects of ionic liquids (ILs) on the lateral diffusion of liver-extract lipids, a blend containing phosphatidylcholine (PC), phosphatidylethanolamine (PE), phosphatidylinositol (PI), and cholesterol. Their study showed that incorporating ILs into liver-extract lipid membranes enhances lateral diffusion, aligning with results from studies on single-component lipid membranes. Similarly, Paterno et al.[126] investigated the impacts of an intramembrane photo-actuator, ZIAPIN2, and its vehicle solvent, dimethyl sulfoxide (DMSO), on the dynamics of both a model membrane (POPC ULV) and intact Human Embryonic Kidney (HEK) cells. They found that while DMSO restricted membrane dynamics, ZIAPIN2 increased them in both model and cellular membranes. These studies underscore the importance of extending research into more complex membrane systems to understand diffusion in realistic biological contexts.

As our capacity to analyze these intricate systems improves, we can expect deeper insights into the mechanisms driving lipid diffusion and how they are influenced by the surrounding environment. This will pave the way for new discoveries in cellular biophysics and membrane dynamics, with potential applications in understanding cell function, disease mechanisms, and targeted drug delivery.

ACKNOWLEDGEMENTS

We are grateful to our collaborators who contributed to work discussed in this manuscript.